# Stochastic Safety Limits and Scale-Dependent Power Fluctuations in Nuclear Reactors: A Critical Scaling Approach

**V. V. Ryazanov**

Institute for Nuclear Research, pr. Nauki, 47 Kiev, Ukraine, e-mail: vryazan19@gmail.com
ORCID ID: 0000-0002-5308-3212

Applying boundary functionals of random risk processes to various physical problems makes it possible to determine many important characteristics of these problems. For example, a special case of boundary functionals is the time to first reach a level, which is widely and successfully applied to a variety of problems. We consider the application of boundary functionals to solving nuclear safety problems. In situations such as reactor startup, as well as for certain types of reactors, neutron behavior changes. Neutron clustering begins to play an important role, and the distributions characterizing neutron behavior change. The normal Gaussian distribution is replaced by stable limiting, distributions to which the sums of random variables converge. Boundary functionals allow us to accurately calculate the statistics of random events, determine the behavior of reactor power peaks, the probabilities of catastrophic power surges, and other quantities important for reactor safety, providing a mathematical bridge between the abstract theory of directed percolation and engineering calculations of protection parameters. This article examines the first-passage time to reach a certain level.

## 1. Introduction

A number of works [1-9] are devoted to the study of stochastic phenomena in the neutron system of a nuclear reactor, namely neutron noise. These include fundamental works on the stochastic nature of neutron fields and experiments confirming this nature. Works [1], [3-9] consider bubble boiling and rod oscillations as sources of information on the state of the core and a basis for describing noise, which can be expanded by applying Levy statistics for strong fluctuations. The time before the onset of instability is also estimated. Works [1], [3-9] are a "bridge" from the microphysics of fission to macroscopic stochastics, which in Ref. [10] is described using the Lundberg equation. In Ref. [7], direct measurements of noise in the core were carried out. The slope of the spectrum in the low-frequency region corresponds to the parameter $a$ in (9) used below for developed bubble boiling. This paper assumes the probabilistic nature of processes in a nuclear reactor and uses stochastic methods to study them, as done in Ref. [10].

The classical theory of neutron noise, developed in Refs. [1, 3-9], convincingly demonstrates that fluctuations in the environmental parameters of power reactors (primarily due to boiling) exceed the intrinsic noise of the fission process by orders of magnitude. Experimental comparison of models based on inverse Kolmogorov equations with spectral analysis data confirmed the presence of long-period correlations. Our study expands this approach, moving from a local description of noise to a global analysis of temporary safety margins at the critical point of directed percolation.

Article [10] delves into an array of boundary functionals within random risk processes, including key metrics like first-passage time, function extremums, moments of return, surplus prior to ruin, the time distribution in the lower half-plane, and the duration spent above a specific threshold $u$, among others. These functionals find applications in various scenarios, such as analyzing unicyclic networks influenced by affinity $A$, asymmetric random walks, nonlinear diffusion phenomena, multi-particle reversible target-binding kinetics, two-level system models, Brownian motion, and general diffusion processes. Due to their expansive mathematical foundation, these constructs lend themselves to applications across numerous domains without significant constraints. The objective of this paper is to investigate how first-passage time can be used to improve safety in nuclear power plant operations. The possibilities of using boundary functionals in nuclear safety are indicated in Ref. [11].

A special case and example of boundary functionals is the first-passage time (FPT) functional. This concept is widely used in scientific research and finds application in physics, chemistry, biology, economics, and other fields Refs. [12, 13]. In the context of nuclear safety, FPT serves as a highly efficient and mathematically robust tool for describing the moment a certain dangerous threshold, such as a dangerous power level, is reached. Research into FPT and other boundary functionals may be particularly significant for non-standard reactor behavior, when the probabilities of catastrophic events increase. In Refs. [14–16], it is shown that in specific scenarios, such as reactor startup or accident analysis (e.g., core failure), or for certain reactor types—such as molten salt reactors (MSRs), high-temperature gas-cooled reactors (HTGRs), and pulverized fuel reactors—methods suitable for studying multifractal systems with anomalous diffusion, such as directed percolation (DP), offer a suitable descriptive framework. In systems driven by directed percolation, reaching a critical dangerous power level ($\Phi_{crit}$) corresponds to the FPT problem in the DP model. FPT involves determining the moment when a branching process first crosses a threshold. Within this model, the number of offspring generated by each node (neutron) is treated as a random variable, denoted k, that obeys a given probability distribution P(k). Moreover, the distribution P(k) obeys a robust power law, allowing for a deeper understanding of the dynamics of these processes Refs. [14-16]. The limiting distribution for the convergence of sums of independent random variables is not only the Gaussian normal distribution (this is a special case), but also the class of infinitely divisible distributions and their subclass, stable distributions. The mathematical description of stable power-law distributions and their relationship to the canonical form of the characteristic function of the Lévy-Khinchin type are investigated in Ref. [17]. Directed percolation is studied in detail and comprehensively in Refs. [18-22].

Within the classical Gaussian paradigm, the distribution of FPT is characterized by a sharp peak around its mean. The probability of the reactor experiencing dangerous power surges is negligible, obeying an exponentially small probability. In contrast, for the DP model with $a$=2 in (9), the influence of Lévy flights and neutron clustering fundamentally alters the dynamics. In particular, the FPT distribution exhibits a pronounced heavy tail in the short time range. This change implies a significant probability that the system can reach its critical, dangerous limit much faster than the mean time estimate suggests. This phenomenon, known as the "early ignition effect," highlights the increased risk associated with such stochastic behavior.

Why is this important for deviations from normal reactor operation? For the vast majority of operating reactors (including WWER, RBMK, and PWR), the distribution of the effective number of progenies is narrow and has a finite variance, for example, the Gaussian distribution and its modification for the number of neutrons produced in a single fission event—the Terrell distribution (exponential decay of tails). A power-law distribution is an "exotic" state that occurs only under specific physical conditions. In Refs. [14-16], the distribution considered is not for the number of neutrons emitted in a single fission event, as in the Terrell distribution, but for the total number of neutrons generated across all subsequent generations by a single "parent" particle. In a conventional reactor, the environment is "cramped" for neutrons. The moderator in WWER is water. Neutrons constantly collide with hydrogen. Their trajectory is a tangled ball (Brownian motion). The central limit theorem is valid. Due to the huge number of small random collisions, the overall statistics of the system tends to a normal (Gaussian) distribution. In such an environment, the probability that one fission chain will suddenly become gigantic (for example, $10^6$ times larger than the average) falls exponentially. This is the "Gaussian truncation" of the tails. The power-law form $P(k)=k^{-a}$ (9) arises under conditions of strong correlation and anisotropy, characteristic of directed percolation. This is possible in cases such as: A). The critical point (critical opalescence). Only at the point $k_{eff}=1$ itself (or in an infinitely small vicinity) does the distribution of chain lengths become a power-law. This is a universal property of all systems undergoing a phase transition and is true for all reactors. But as soon as the reactor goes supercritical (runaway), nonlinear feedbacks (Doppler) immediately "truncate" this tail, returning the system to a predictable form. B). Strong spatial inhomogeneity ("Lévy glass"). This occurs for a medium consisting of empty channels and dense fuel blocks (e.g., a destroyed core or specific experimental assemblies). A neutron in such a medium can fly a huge distance without collisions ("Lévy flight"). In such a medium, "descendants" are born not as a compact cloud, but as a fractal structure. Here, the geometry of the medium imposes a power law on the distribution of the descendants. C). Small number of neutrons (small sample statistics). In startup modes, when there are only a few hundred neutrons in the core, averaging does not work. Individual "lucky" fission chains can dominate. Researchers (as in articles [14-16]) show that under these conditions, the dynamics are more similar to directed percolation than to diffusion. In Ref. [3] it is shown that the distribution of the mean free path or number of secondary neutrons obeys a power law in the presence of strong spatio-temporal correlations caused by the motion of the components of the active zone.

If neutrons arrive in clusters (as predicted by DP), the probability of pulse pile-up increases sharply. This leads to an underestimation of the actual power during cluster bursts. An engineer might see 1% of the power on the control panel, but in the local cluster it could be as high as 10%.

Reactor startup is a dangerous moment. At the minimum control level (MCL), feedback loops are not yet operational (the reactor is "cold," with the Doppler effect and thermal coefficient barely perceptible). The system lacks "brakes" (nonlinearities) that would cut off the heavy tails of the distribution. If a "Lévy flight" (an abnormal reactivity spike due to a rod movement or fluctuation) occurs at this point, only physical burnup or an emergency shutdown can stop it. During startup, the reactor is not a homogeneous body, but a collection of "flaring" clusters. The probability of a local runaway at the MCL is higher than classical predictions due to directed percolation effects. Diagnostics require analysis of the higher moments of the distribution (not only the mean and variance, but also the excess) to understand how "heavy" the tails of neutron fluctuations are.

No less effectively than FPT, boundary functionals such as the extrema of a random process on a given time interval, the value of a jump above a certain level, the time a process spends above a level, and other boundary functionals can be used for nuclear safety problems.

The article is organized as follows. Section 2 provides some mathematical definitions of the quantities used in the article. Section 3 examines the power-law distribution for the number of effective progenies. Section 4 examines the components of the Lundberg equation. Section 5 addresses the technical issue of normalizing the characteristic functions of stable distributions. Section 6 examines the influence of solutions to the Lundberg equation on the average time to reach hazardous levels. Section 7 presents the results of calculating this time for different representations of the characteristic functions of stable distributions and various approximations. Section 8 discusses the results obtained. Section 9 offers concluding remarks.

**List of acronyms**: FPT = first-passage time. DP = directed percolation. MSR = molten salt reactor. HTGR = high-temperature gas-cooled reactor. NPP = nuclear power plant. WWER = water water energetic reactor., PWR = pressurized water reactor. BWR = boiling water reactor. MCL = minimum control level. EP = emergency protection. IMS = information management system. PSA = probabilistic safety analysis. CF = characteristic function. DFA = detrended fluctuation analysis. I&C = Instrumentation and Control.

**Nomenclature**

- ***a*** — structural indicator of the distribution of offspring ($P(k) \sim k^{-a}$). The main parameter of "tail heaviness".
- **α=*a*-1** — stability index of Levy's law (from Zolotarev's form [17]). For directed percolation $\alpha \approx 1$.
- **s** — Laplace transform parameter over time [$T^{-1}$]. Conjugate to the threshold reaching time t.
- **k** — Laplace transform parameter by power (Lundberg equation root) [$\Phi^{-1}$].
- **v** — effective drift velocity (reactivity) [Φ/T] or [1/T].
- **σ** — intensity of stochastic divisions (percolation parameter).
- **λ≈1/L** — exponential truncation parameter (fuel assemblies (FA) geometry) [$L^{-1}$].
- **C(*a*),** $C(a) = |\Gamma(2-a)\cos((a-1)\pi/2)|$ — coefficient obtained from the Γ-functions.

## 2. Notations and definitions

Let us define FPT as the time it takes for the process ξ(t) to reach a positive level x>0 Ref. [10]. Below, only positive levels will be considered, and there will be no index +;

$\tau^+(x) = \inf\{t : \xi(t) > x\},\ \ x > 0$ is the moment of the first exit for the level $x$>0; (1)

In our case, the definition used below is:

$$T_{FPT}=\inf\{t>0: \Phi(t)\geq\Phi_{crit}\}, \tag{2}$$

where Φ(t) is the neutron flux or reactor power, considered as a random process, $\Phi_{crit}$ is some critical value of this quantity.

A process $\{\xi(t),\ 0\leq t\leq T\}$ is called a process with independent increments if for an arbitrary $n\geq1$ and for all $\{t_k\}_{k=\overline{0,n}}$, $0\leq t_0<t_1<...<t_n\leq T$, the values $\xi(t_0)$ and increments $\xi(t_0)$, $\xi(t_1)$ - $\xi(t_0),\ldots,\xi(t_n)-\xi(t_{n-1})$ are independent random variables. Note that in Ref. [23] the assumption of independence of increments is replaced by the renewal principle.

A process with independent increments is called homogeneous if the distribution of increments $\xi(t)-\xi(s)$ depends only on *t-s* when $s\leq t$. If $\xi(t)$ is a homogeneous process with independent increments, then we will assume that $0\leq t<\infty$, and it is often assumed that $\xi(0)=0$.

Let *X* be a random variable with cumulative distribution function $F_X$. The moment generating function (*MGF*) of *X* (or $F_X$), denoted by $M_X(t)$, is:

$$M_X(t)=\boldsymbol{E}[e^{tX}] \tag{3}$$

provided this expectation exists for *t* in some open neighborhood of 0. That is, there is an $h>0$ such that for all *t* in $-h<t<h$, $\boldsymbol{E}[e^{tX}]$ exists. If the expectation does not exist in an open neighborhood of 0, we say that the moment generating function does not exist.

If *X* is a continuous random variable, the moment-generating function's (MGF) definition expands (by the law of the unconscious statistician) to:

$$M_X(t)=\boldsymbol{E}[e^{tX}]=\int_{-\infty}^{\infty}e^{tx}f_X(x)dx, \tag{4}$$

where $f_X(x)$ is probability density function.

If $\xi(t)$ is a homogeneous stochastically continuous process with independent increments in $R^m$, then its characteristic function takes the form (a Lévy-Khinchin type representation):

$$\boldsymbol{E}e^{i(z,\ \xi(t))}=\exp\{t[\,i\,(z,a)-\frac{1}{2}(Bz,z)+\int_{|x|\leq1}(e^{i\,(z,x)}-1-(z,\ x))\Pi(dx)+\int_{|x|>1}(e^{i\,(z,x)}-1)\Pi(dx)]\}, \tag{5}$$

where $a\in R^m$; $B$ is a symmetric non-negative operator in $R^m$; $\Pi$ is a measure in $R^m$, for which $\int\frac{(x,x)}{1+(x,x)}\Pi(dx)<\infty$, and $\Pi(\{0\})=0$.

Thus, the first assumption about the homogeneity of the random process contains a limitation.

The characteristic function (CF) of a random variable $\xi$ with a distribution function $F(x)=P\{\xi<x\}$ is referred to as a complex-valued function:

$$\varphi(t)=\boldsymbol{E}e^{it\xi(t)}=\int_{-\infty}^{\infty}e^{itx}dF(x). \tag{6}$$

The characteristic function of a homogeneous process $\xi(t),\ t\geq0$ is determined in the theory of random processes (for $\xi(0)=0$) (works [10, 12, 17, 18]) by the relation:

$$\boldsymbol{E}e^{i\alpha\xi(t)}:=\int_{-\infty}^{\infty}e^{i\alpha x}dF(x)=e^{t\Psi(\alpha)},\ t\geq0, \tag{7}$$

where $F(x)=P(\xi<x)$ is the distribution function of a random process $\xi(t),\ t\geq0$, and the function $\psi(\alpha)$ represents the scaled cumulant generating function (*SCGF*) of the process $\xi(t),\ t\geq0$. The *SCGF* do not depend on *t* and characterize all finite-dimensional distributions of the process.

If for a process $\xi(t),\ t\geq0$, the function $\psi(\alpha)$ at $i\alpha=r$ is equal to *s*, then we obtain the equation:

$$\Psi(\alpha)\big|_{i\alpha=r}:=k(r)=s,\quad\pm\mathrm{Re}\,r\geq0. \tag{8}$$

This equation, in risk theory, is referred to as the fundamental Lundberg equation.

## 3. FPT and number of effective progenies

In our model, the number of offspring generated by each node (neutron) is treated as a random variable, denoted k, that follows a given probability distribution P(k). We assume that in some situations, the distribution P(k) follows a stable power law:

$$P(k)\sim k^{-a}. \quad (9)$$

Such a dependence was obtained, for example, in Refs. [24, 25]. In Refs. [14-16] it was shown that the branching random walk of neutrons in the presence of feedback mechanisms (for example, the Doppler effect) can belong to the universality class of directed percolation. In these models, the critical decay or growth of clusters is described by power laws. In Ref. [14], the distribution of the total number of descendants of a single neutron (the cluster size in space-time) is analyzed. In the critical state $k_{eff}\approx1$, this distribution takes on the power-law form (9) $P(k)\sim k^{-a}$, where the exponent $a$ is determined by the critical indices of percolation theory. The work [14] proved that the critical point of a nuclear reactor corresponds to a phase transition of directed percolation. The authors (including Ref. [14]) consider not the fission of a single nucleus, but the effective number of progenies in a heterogeneous medium. A single neutron can trigger a chain of fissions in a local fuel cluster ("subcritical breeding"). If the medium is fractal or is near the critical point, then the effective number of neutrons emitted from such a "super-node" can have a very wide spread. It is this "effective" quantity that can behave as a power law on the scale of a macroscopic lattice. The article [16] proves that in a critical reactor, neutrons are distributed not as a rarefied gas (diffusion), but as patchy structures. The authors show that, due to the branching process (fission), neutrons have common "ancestors", which forces them to stick together in groups. This is a direct manifestation of directed percolation in space-time: fissions chains form tree-like structures, which become fractal near $k_{eff}=1$.

The power-law distribution is valid for a critical branching process near a bifurcation point. In the context of a nuclear reactor, it describes the probability distribution of the total number of neutrons produced by a single initial neutron chain (of the total chain size n) when the reactor is exactly at the critical state $k_{eff}=1$). In general, this is not the distribution of the number of immediate descendants of a single neutron k, which is usually approximated by a Poisson distribution. It is the distribution over the total number of neutrons in the chain, including all generations up to its extinction. Under special conditions (very small, "pulsating" assemblies, strongly absorbing environments where spatial fluctuations are critically important), deviations may be observed, and a more appropriate model may be the DP at d=3 or d=2 (for planar assemblies). The neutron lifetime is short, but it manages to diffuse a significant distance before fission. This effectively "washes out" spatial correlations, increasing the effective dimensionality of the system to the critical value $d_c=4$, where the mean-field exponents (including 3/2) apply. If spatial fluctuations in a 3D reactor are taken into account (the real physical case), the exponent $a$ is corrected by the critical DP indices: $a=1+\beta/(\beta+\nu_{II})$, where for 3D directed percolation: $\beta\approx0.81$ (density index), $\nu_{II}\approx1.90$) (correlation time index). This yields a value of $a\approx1.3$. In addition to the critical point, the transition to a power-law regime occurs in such specific cases as the action of feedback effects: when taking into account nonlinear effects (for example, temperature feedback), which transform the system into a self-organizing critical environment; strong spatial heterogeneity, when the structure of the active zone or neutron fluxes acquires fractal properties. In the article [14], it is substantiated that the "effective" number of descendants is not simply the number of fissions in all generations, but the total contribution of the branching process, limited by "leakage" or absorption.

The power law arises because, near criticality, the probability of long chains decreases slowly: instead of exponential decay $\exp[-k/<k>]$, we get algebraic decay $k^{-a}$. In a reactor, this explains "neutron clustering"—zones of anomalously high neutron density that appear "out of nowhere" and then disappear, which is impossible within the framework of standard diffusion.

In Refs. [14-16] it is emphasized that if the distribution of neutron ranges is itself "heavy" (for example, due to voids in the geometry), then the process goes into the Lévy-DP regime, where the exponent $a$ begins to depend on the Lévy flight parameter $\alpha$.

Experimental confirmation (Experiments at LANL). The DP theory was tested on a real critical assembly (Planet, Los Alamos) [16]. The authors measured correlations between fissions and found that the spatial structure of the flux fluctuates anomalously. The statistics of these fluctuations are not described by a Poisson (classical) distribution. They exhibit "heavy tails," which directly indicates Lévy statistics and effects close to $a$=2.

In Ref. [14], the dependence $N(x) \sim x^{\beta}$ is written for prompt neutrons in the case of DP. In most physical and statistical systems, where a random variable (for example, the number of nodes or the cluster volume N) is scaled in this way, the probability distribution P(N) also obeys a power law. In Ref. [28], it is shown that in the critical regime, the process trajectories for the total number of neutrons behave like ta, increasing according to a power law. More precisely, the random process behaves as x(t)~t h, where h is a random variable uniformly distributed over the interval (0,1), i.e. P{h<x}=x over the interval (0,1). The "effective number of descendants" k refers to the total number of neutrons generated across all subsequent generations by a single "parent" particle under specific physical conditions. In a conventional WWER reactor, the distribution of these descendants does not follow a power-law pattern. However, a power-law distribution (type (9)) can emerge under conditions characterized by strong correlations and anisotropy, which are hallmarks of directed percolation. This phenomenon typically occurs in the three scenarios noted in the introduction. The power-law tail exponent ($a$=α+1) of the probability density function of the stable distribution for large values of k in (9) decreases according to a power law. Confusion between $a$ and α can arise due to the form of notation of the probability density function of the stable distribution. The stability index (α) determines the properties of the distribution and is included in the characteristic function as $\exp(-|k|^{\alpha})$. It determines the scaling properties (the self-similarity parameter).

If the distribution of the number of effective progeny P(k) obeys a power law $k^{-a}$ (9) with $a$≈2, the system leaves the Gaussian universality class (normal distribution). In a real fission event, it is physically impossible to emit 100 or 1000 neutrons (P(ν) instantly vanishes after ν=8 due to the limited binding energy of the nucleus). In a classical reactor (WWER), the "physical" P(ν) is a narrow Gaussian/Terrell. DP effects with power-law tails are properties not of the nucleus itself, but of the collective behavior of the neutron gas in a complex geometry.

For a distribution function describing the number of descendants or jump lengths, expressed in the form (9) with a power-law tail proportional to $k^{-a}$, the characteristic function $\Phi(t)=E(e^{itX})$ associated with such power-law tails exhibits a non-analytic behavior at zero $\ln\Phi(t) \sim -|t|^{\alpha}$ Ref. [17]. This feature $\ln\Phi(t)$ aligns with the Lévy-Khinchin representation (5) for stable distributions. When transitioning to the continuous limit at the macroscopic scale (e.g., using a differential equation), this term $|t|^{\alpha}$ corresponds to a fractional derivative with respect to space or time Refs. [26, 27

The value $a$=2 is a "turning point" for statistical moments. If $a$≤2, then the mathematical expectation (the average number of offspring) <k> diverges (tends to infinity for an infinite system size). If 2<$a$≤3, then the mean is finite, but the second moment $\langle k^2 \rangle$ diverges. If $a$≈3, then both the mean and the variance are finite. When the variance is finite (a≈3), the classical central limit theorem applies to the sum of such quantities, and the behavior of the system becomes "typical" in many aspects. For $a$≈3 (where variance is finite), the random walk through many steps is described by the ordinary diffusion equation (the normal distribution of coordinates). For a<3 (including the case a=2), Lévy flights occur.

The utilization of FPT functionals in the analysis of directed percolation models facilitates the estimation of probabilities associated with "instantaneous" local burnouts. Moreover, it enables an assessment of the adequacy of protective systems' response times while also incorporating the stochastic characteristics of activation processes. This approach provides a significant advantage over deterministic computational codes such as KORSAR or RELAP, which typically overlook these probabilistic aspects.

Article [10] transitions from analyzing mere average values to exploring the complete statistical characteristics of boundary functionals in stochastic processes, including FPT Ref. [10]. While it delves into the FPT in detail, the article [10] also derives several other boundary functionals, which are equally critical for ensuring nuclear power plant (NPP) safety.

For processes governed by stable Lévy laws (with an index $\alpha = a - 1 \approx 1$), the probability density related to the time it takes to first reach the threshold L exhibits the following asymptotic behavior Refs. [19, 20, 25]:

$$P(t_{FPT}) \sim t^{-(1+\alpha/z)}. \quad (10)$$

In this context, z represents the dynamic critical index associated with directed percolation. The average time to threshold, denoted as $\langle t_{FPT} \rangle$, may theoretically diverge or exhibit an exceptionally large variance. Consequently, the notion of a "mean time to failure" becomes irrelevant in practical terms for safety considerations, as the actual distribution of times required to reach the critical threshold spans multiple orders of magnitude. In traditional kinetics (point kinetics equations), fluctuations are considered Gaussian. However, near criticality, the reactor behaves like a system undergoing a phase transition. In this case, neutron clusters develop as percolation paths, and the time to reach the emergency threshold L obeys this power law, not an exponential one. Fundamental scaling laws for the density of active sites and the survival probability $P(t) \sim t^{-\delta}$ are described in Ref. [20]. Formula (10) for the FPT is a classic result of the theory of processes with "heavy tails." For a critical reactor, the effective step of the process (neutron transport) and the time scale are related through the ratio $\alpha/z$.

If the number of secondary neutrons or the mean free path has heavy tails, the traditional diffusion equation is replaced by a fractional diffusion equation Refs. [26, 27]. The concept of "neutron transport in correlated media" (Lévy flights) is mentioned in Ref. [3]. Fractional calculus is applied in Refs. [26-27]. It was shown that in highly inhomogeneous media or media with a fractal structure (e.g., porous absorbers), neutrons move not as particles in a gas, but as "Lévy jumpers."

This approach can be directly applied to WWER safety assessments, particularly during reactor startup or operation at the Minimum Control Level (MCL). The utilization of the First Passage Time (FPT) functionality addresses three critical aspects: a). Adjustment of Emergency Protection (EP) Setpoints: The EP setpoints, particularly for the rate of increase or period, must be configured carefully. The goal is to filter out fast, transient spikes from the tail of the FPT distribution without triggering false alarms due to normal statistical noise. This ensures that safety mechanisms are activated only under genuine risk conditions. b). Equipment Dead Time: If the FPT distribution exhibits a heavy tail in the region corresponding to short times, there is a potential hazard. It indicates that the system's acceleration time might fall below the total protection response time (which includes the rod drop time and the time required for Information Management System (IMS) logic to process and act). This could compromise the timely deployment of protective measures. c). Probabilistic Safety Analysis (PSA): Instead of relying on a single predefined accident scenario, an FPT distribution is constructed to provide a broader, probabilistic perspective. For safety to be deemed sufficient, the probability integral $P(t_{FPT} < t_{action})$ — where $t_{action}$ represents the system response time — should be negligible. This ensures that there is an exceedingly low likelihood of the system response being outpaced by an accelerating risk. By leveraging these analytical approaches, safety measures for WWER reactors can be optimized and tailored for precise and effective risk mitigation.

The finiteness of the reactor's size (L) imposes a truncation effect on Lévy flights. In the context of first-passage times (FPT), this truncation ensures that, over extended timescales, the distribution will asymptotically transition to an exponential form. Nonetheless, it is important to note that critical thresholds or hazardous conditions typically arise within shorter timescales, during which the impact of truncation remains negligible. Within this high-risk domain characterized by rapid accelerations, the reactor exhibits dynamics consistent with those of a "pure" directed percolation system, retaining all the associated anomalies inherent to such systems.

Let us write the formula for the truncated Levy distribution. Unlike the pure power law $P(k) \sim k^{-(1+\alpha)}$ (9), $a=1+\alpha$, which leads to divergence of moments (infinite variance), the truncated distribution introduces an exponential damping factor related to the reactor geometry (L) and the physics of capture:

$$P(k) \approx N \cdot k^{-(1+\alpha)} \cdot \exp(-\lambda k), \quad a = 1+\alpha, \quad (11)$$

where $\alpha$ is the stability index (for the case $a=2$ we have $\alpha=1$), $\lambda \sim 1/L$ is the truncation parameter determined by the characteristic size of the active zone or the path length to capture, N is the normalization coefficient.

The connection with the Lévy-Khinchin formula here manifests itself in the modification of the logarithm of the characteristic function lnΦ(q). Instead of pure $|q|^{\alpha}$, which yields DP correlations, we obtain:

$$\ln\Phi(q) \sim const((\lambda^2+q^2)^{\alpha/2}\cos(\alpha\arctan|q|\lambda)-\lambda^{\alpha}) . \tag{12}$$

Estimates show that the most dangerous situation for a WWER reactor is collective excitation within a single fuel assembly. Individual fluctuations in fuel rods are safe due to high λ. Global fluctuations of the entire reactor are suppressed by the Doppler effect and the reactor's enormous size. The "vulnerability window" is located precisely at the scale of a single fuel assembly (20-30 cm), where directed percolation is already effective and geometric truncation is still weak. The correlation length of noise is examined in Rev. [8]. Upon reaching the instability threshold, the correlation scale becomes comparable to the fuel assembly size (L=25 cm).

It can be concluded that the VVER geometric structure creates a hierarchy of stochastic risks. While percolation effects are suppressed at the microlevel (fuel rods) (λ≈1, λ≈1/L), at the mesolevel (fuel assembly) (λ≈0.04, L=25), the statistics of directed percolation with a power-law exponent *a*=2 can lead to abnormally fast local transient processes, requiring special attention when setting up core monitoring. A factor of L=25 maximizes the effective rate of stochastic runaway, reducing the time <T> to an accident by tens of percent.

The truncation parameter λ (or scale L) is most dangerous at intermediate intervals (the fuel assembly scale). At *a*=3, the system is local and geometry-insensitive. A neutron "doesn't see" the geometry of the entire assembly; it sees only its nearest neighbors. The system becomes local. At *a*=2, a resonance of stochastics and size occurs at the fuel assembly scale. However, at the scale of the entire core, global feedbacks suppress percolation effects, forcing the system back into the region of effective Gaussian distributions.

## 4. Lundberg equation

Expressions for the moments of the FPT value are written in Ref. [10]. The relations for continuous and semi-continuous (top and bottom) processes differ in the pre-exponential factors before the exponential; an expression for the MGF $T(s,x)=\boldsymbol{E}[e^{-s\tau^+(x)},\tau^+(x)<\infty]$ of the form was obtained:

$$T(s,x)=\boldsymbol{E}[e^{-s\tau^+(x)},\tau^+(x)<\infty]=e^{-\rho_+(s)x},\ \ x>0\,, \tag{13}$$

$$\boldsymbol{E}[\tau^+(x),s]=-\frac{\partial\ln\boldsymbol{E}[e^{-s\tau^+(x)},\,\tau^+(x)<\infty]}{\partial s}=\frac{\boldsymbol{E}[\tau^+(x)e^{-s\tau^+(x)},\,\tau^+(x)<\infty]}{\boldsymbol{E}[e^{-s\tau^+(x)},\,\tau^+(x)<\infty]}=x\frac{\partial\rho_+(s)}{\partial s}\,.$$

The calculations carried out in [10] show that there is practically no difference between these expressions.

In Ref. [10], the expressions for the boundary functionals (13) include the roots of the Lundberg equation (8), (14). The Lundberg equation (k(r) from (8)) $k(r)_{r=k}$=G(k)=s is the search for such a value k(s)=ρ+(s), at which the process generator (SCGF (7)-(8)) balances the Laplace transform parameter over time s. The Lundberg equation (14) translates events at the micro level into seconds of time before the accident (macro level);

$$G(k)=s, \tag{14}$$

B (14) G(k) is the logarithm of the characteristic function (more precisely, the moment-generating function's (MGF) (4) M(k)=E[$e^{kX}$], obtained from the characteristic function (6) by analytical continuation and replacement iq→k), and s is the Laplace transform parameter for time. For the case *a*=2 (critical DP), the Lévy-Khinchin integral is calculated analytically, leading to a logarithmic structure of the logarithm of the MGF $\Psi(q)=-\sigma|q|(1+i\beta(2/\pi)\operatorname{sgn}(q)\ln|q|)+i\mu q$ Ref. [17]. This explicit form, corresponding to the critical α=1, transforms the Lundberg equation (14) [10] into a logarithmically "clamped" asymptotic, explaining the anomalously high-power peaks. The Lundberg equation for the case of stable Lévy distributions (*a*≈2, α≈1) is transcendental. Its solution k(s) (notation $\rho_+$(s) in (13) Ref. [10]) determines the behavior of the boundary functionals Ref. [10] (FPT, maximum of a random process, etc.). In the case of directed heavy-tailed percolation, we do not have a simple quadratic solution, as in ordinary diffusion ($Dk(s)^2=s\rightarrow k(s)=\sqrt{s/D}$).

In Ref. [17] it is noted that there are many different forms of recording the characteristic functions of stable laws, special cases of the canonical form (5). Five of them are given in Ref. [17]. In all five (as in other forms) there is a power dependence on the argument of the characteristic function, corresponding to the power dependence of the stable distribution function of the form (9).

For a stable law with index $\alpha\in(0.2]$, $a\in(1.3]$, the logarithm of the characteristic function in canonical form (A) Ref. [17] looks like this (17):

$$\ln\Phi(q)=i\gamma_1 q-\sigma|q|^{\alpha-1}[|q|-\frac{iq}{|q|^{\alpha-1}}\omega(q,\alpha)],\ \ \omega(q,\alpha)=|q|^{\alpha-1}\beta tg(\frac{\pi\alpha}{2})\ (\alpha\neq1),\ \omega(q,\alpha)=-\beta\frac{2}{\pi}\log|q|\ (\alpha=1),$$

where $\gamma_1$ is the shift parameter (drift), σ is the scale parameter, $\beta\in[-1,1]$ is the asymmetry parameter; for maximum asymmetry β=1, which corresponds only to positive power surges (Appendix B).

In problems on boundary functionals (FPT, maximum, etc.), which are considered in [10], we move from characteristic functions (CF) (6) (Fourier transform with argument iq, CF (6)) to moment generating functions (4) (Laplace transform, parameter p or k). A change of variable is made. In the Lundberg equation, the argument k is often considered as a real parameter of the Laplace transform. This is equivalent to substituting $q\rightarrow -ik$ into the Fourier formula.

Asymmetry and positivity are important. We model the number of neutrons or the power ($\Phi\geq0$), and the process is one-sided (strictly positive jumps). For such processes in Lévy's theory, the asymmetry parameter β=1 (maximum asymmetry). The imaginary parts in the expression from Ref. [17] are canceled or transferred to the real domain under certain transformations, describing a purely decaying or growing process.

After the transition to the real form G(k) for strictly positive stable processes with index $\alpha<1$, the logarithm of the Laplace transform has a very simple form without imaginary units:

$$G(k)=\ln E[e^{kX}]=-\sigma k^{\alpha}. \tag{15}$$

This form is used in physics applications, as it describes the probability of chain survival. For the case $a$=2 ($\alpha\rightarrow1$), if the process is symmetric, $G(k)\sim-|k|^{\alpha}$ (there is no imaginary part). If there is directionality (drift v), the term vk appears.

In the articles [10] (and in the theory of branching processes [28]), G(k) in (15) is understood to be a cumulant function: $G(k)=\ln\sum P(n)e^{-kn}$ . For power-law distributions, $P(n)\sim n^{-(1+\alpha)}$ (9) this integral (or sum) in the limit of small k gives precisely the power-law function $k^{\alpha}$.

In Ref. [17], the characteristic function (6) of the stable law $\Phi(t)=E[e^{itX}]$ has the form (form (C) in Ref. [17]):

$$\ln\Phi(t)=-\sigma|-t|^{\alpha}\exp[-i\frac{\pi\alpha\theta}{2}sign(t)],\ \ 0<\alpha\leq2,\ \ |\theta|\leq\theta_{\alpha}=\min(1,\frac{2}{\alpha}-1). \tag{16}$$

In the Lundberg equation in Ref. [10] (and in Ref. [28]), we work with the MGF $M(k)=E[e^{kX}]$ (4); at t=-ik:

$$\ln M(k)=-\sigma|-ik|^{\alpha}\exp[-i\frac{\pi\alpha\theta}{2}sign(-ik)]=-\sigma|k|^{\alpha}\exp[-\frac{\pi\alpha\theta}{2}sign(k)].$$

Since $|-ik|=k$ , and sign(-ik)=-i (in the complex plane), after transforming the phase factors $\exp(i\pi\alpha/2)$ we obtain a real result for $G(k)=\ln M(k)$.

To ensure that the formula corresponds to the canonical form for one-way processes, the theory of stable laws uses the cosine relationship. For a strict transition (taking into account the normalization of σ in Ref. [17]): $G(k)=\ln M(k)\sim C(a)k^{\alpha}$. For k<0, for the growth of the system, this becomes $-\sigma_{eff}\,|k|^{\alpha}$.

The imaginary unit *i* in the Fourier transform describes the phase shift. By transitioning to the Laplace transform (t=-ik), we "rotate" the solution from the oscillation (wave) region to the exponential growth/decay region. Since our process (power) physically cannot be negative, all imaginary components must cancel out during this rotation. If *i* remained, this would mean that the probability of reaching the level is a complex number, which is absurd.

As $\alpha \to 1$ (case $a$=2) the ratio $C(a) = |\Gamma(2-a)\cos((a-1)\pi/2)|$ turns into an uncertainty $\infty \cdot 0$. When expanded through Γ-functions (as in the more general Lévy-Khinchin form [17]) this coefficient gives π/2, and $G(k)=vk-\sigma(\pi/2)|k|^{\alpha}$.

The Lundberg equation is often written for negative values of the argument in the exponential (exp[-kX]), since this guarantees convergence for positive values of X. If we define $G(k)=\ln E[e^{kX}]$, then for a stable process with one-sided jumps at k<0 (exponential decay): G(k) behaves like $C(-k)^{\alpha}$. The sign in front of σ depends on whether we consider the "incoming" of neutrons (multiplication) or the "outgoing" (absorption/diffusion).

## 5. On cosine normalization (connection with Riesz derivatives)

Normalization via the cosine arises in the definition of the fractional Laplacian (or Riesz-Feller derivative). In the theory of fractional calculus, the relationship between the operator da/dxa and its image in k-space is written as follows: $F\{d^a/dx^a\} \sim -|k|^a \cos(\pi a/2)$. To eliminate this cosine in the final equations and work with pure $|k|^a$, physicists often "transfer" it to the definition of the coefficient σ.

The factor Γ(2-$a$)cos(π($a$-1)/2) is essentially a "bridge." For $a \to 2$ (diffusion), the cosine vanishes, and the Γ-function tends to infinity. Their product yields π/2. This is exactly the value required for the fractional derivative of order 2 to transform into the ordinary second derivative $d^2/dx^2$. The Γ-function is the result of direct integration of the jump measure. Substitution in form (A) yields a real expression with a tangent. The logarithm at $a$=2 (17) is the only way to remove the discontinuity of this tangent at critical points while maintaining analyticity.

The calculation formula is a regularized form of (A) Ref. [17]. It is physically correct, since in the limit it gives classical diffusion and at the same time takes into account Lévy's "shoot-throughs" through the Γ-function. The calculation combines classical reactor theory (the diffusion approximation) with modern anomalous transport theory. The minimum time to reach $a$=1 is a "fundamental resonance" between the fuel assembly geometry and the percolation type.

## 6. Solution of Lundberg equation and Regularized Response Function

The generator G(k) in equation (14) is the logarithm of the MGF of form (4). For asymmetric processes (power growth), in the case of using form (A) Ref. [17], it is written as:

$$G(k)=vk+\sigma C(a)|k|^{a-1}[1+\beta\cdot\mathrm{sign}(k)\tan(\pi(a-1)/2)],\ a\neq 2;\quad G(k)=vk+\sigma C(a)|k|[k^{a-2}-\beta\cdot 2\log|k|/\pi],\ a=2, \quad (17)$$

$$\beta=0,\quad G(k) = vk + \sigma C(a)|k|^{a-1}, \qquad C(a) = |\Gamma(2-a)\cos((a-1)\pi/2)|,$$

where v is the deterministic drift (effective reactivity), and the second term is the stochastic contribution $V_{stoch}=\sigma\cdot C(a)\cdot k^{a-1}$, β=0; in the general case $V_{stoch}(a, \beta, L)=\sigma\cdot\mathscr{R}(a, L)\cdot\Psi(a, \beta)$, where $\mathscr{R}(a, L)$ is given in (20), and $\Psi(a, \beta)=\Gamma(2-a)\,[|\cos(\pi(a-1)/2)|+\beta\sin(\pi(a-1)/2)]$ (Appendix B).

As obtained in expression (13), the average time <T> is related to the derivative of the generator G`(k); from (14) we find: $G'(k)$dk(s)/ds=1, dk(s)/ds=1/$G'(k)$,

$$\langle T\rangle = \Phi_{crit}\,\mathrm{dk(s)/ds} = \Phi_{crit}/G'(k). \quad (18)$$

If at β=0, $G(k)=vk+\sigma C(a)|k|^{a-1}$ (17), then its derivative:

$$G'(k) = v + \sigma C(a)(a-1)|k|^{a-2}. \quad (19)$$

In a more general case with $\Psi(a, \beta)$, and where not $\mathscr{R}$(a,L) appears, but $L^{\delta(a)}$:

$$\langle T\rangle(a)=\Phi_{crit}/(v+V_{st}), \qquad V_{st} = \sigma\cdot L^{\delta(a)}\,[C(a)+\beta\Gamma(2-a)\cdot\sin(\pi(a-1)/2)].$$

In statistical mechanics, the Laplace transform (the generator G(k)) is closely related to the generating function of cumulants: $\psi(k)=\ln\langle e^{kX}\rangle$.

In trajectory thermodynamics Refs. [29-30], Ψ(k)=G(k) plays the role of dynamic free energy, which corresponds to the topological pressure of dynamic system theory. In equilibrium thermodynamics, k is the conjugate field (like the inverse temperature β). In our problem, k is the parameter of the control field, which "selects" from all possible reactor trajectories those that lead to an emergency release. It is possible to find critical values of the "field" k and the index *a*, at which the system undergoes a dynamic phase transition from a stable state to the state of a "ruined" (emergency) system.

We consider the logarithm of MGF as the potential of 'dynamical free energy' in the s-ensemble of trajectories Refs. [29–30]. The singularity of this potential at *a*=2 is the mathematical reason why a 30% increase in risk (from (17)) becomes inevitable.

At the point *a*=1, a singularity of the potential Ψ(k)→∝ itself occurs (structural catastrophe), the potential ceases to be smooth at k=0. In the theory of stable distributions Ref. [17], the condition *a*>1 (α>0) is valid. The neutron population can violate this condition. However, in finite systems, the equality k=0 is not satisfied. For *a*>1 (short-range regime), the system behaves as in ordinary directed percolation (DP). The long-range action decays quickly enough not to affect the critical exponents. For a~1 (long-range regime), "Lévy flights" become dominant. For *a*=2, the potential remains finite, but its first derivative becomes singular, which marks a change in the regime from short-range to a regime with dominant long-range action ("Lévy flights"), changing the universality class of the system. The system passes to a new regime, where the critical exponents begin to depend on the parameter *a*. A first-derivative singularity in this context denotes a jump or break in thermodynamic quantities or transport coefficients. This is a classic sign of a phase transition (often second-order or higher, according to Ehrenfest's classification), where the system changes its global response to small perturbations. This is a cross-phase transition in directed percolation models.

At the point of directed percolation (*a*=2), the following occurs: a change in the sign of the curvature: up to *a*=2, the free energy Ψ(k) is convex, which corresponds to the central limit theorem (stability). At the point *a*=2, the curvature (analogous to specific heat) becomes infinite. Condensation of trajectories: neutron trajectories begin to "condense" into narrow, long clusters. This is analogous to the "gas-liquid" phase transition, where the "liquid" is the continuous chain of fission events throughout the entire fuel assembly. Logarithmic contribution: it is at this point that, due to critical fluctuations, the correlation length ξ becomes equal to L. The contribution to the denominator becomes σlnL. The work [14] proved that the critical point of a nuclear reactor corresponds to the phase transition of directed percolation.

The free energy of trajectories Ψ(k)=G(k) exhibits different types of singularities: at *a*=3, a singularity of the second derivative Ψ``(k)→∝ (second-order phase transition) occurs. At *a*=2, a singularity of the first derivative of the potential Ψ`(k)→∝ occurs. A singularity of the first derivative is always a sharp release of energy or a sharp change in density.

The risks associated with underestimating local criticality are explored in Refs. [14-16] on "neutron clustering" (particle clustering). Studies of stochastic environments discuss the effect of unexpected criticality during density fluctuations, whereby a difference of 1–2% on average can translate into significant deviations in local fluctuations.

There are two possible ways to write the G(k) function. The difference in signs (plus or minus) and absolute values depends on the analytical continuation to the complex plane. The minus sign at β=0 ( $\nu k - \sigma C(a)|k|^{a-1}$ (17)) is typically used in diffusion problems (the Fokker-Planck fractal equation), where stochastics "blur" the concentration (reduce it at the center). The plus sign ( $\nu k + \sigma C(a)|k|^{a-1}$ ) is used in risk theory problems Ref. [10]. Here, stochastic jumps (Lévy flights) "help" the system reach the threshold faster. Since we are looking for the root of the G(k)=s equation, where s is the absorption rate, the plus sign physically means that random power surges shorten the time <T>. We are considering the most dangerous case.

We use a form that emphasizes the stability index: $G(k) = \nu k + \sigma C(a)|k|^{a-1}$. Here, k is taken positive, since we are seeking the root in the right half-plane of the Laplace transform. Stochastic "flights" in our

model operate in the same direction as positive reactivity—they accelerate the attainment of the critical threshold. The physical meaning of the choice of the "plus" sign is simple: v is the average velocity of the upward power "drift" (the deterministic part). The second term is the contribution of stochastic "Lévy flights." Since the "flights" (power surges) are directed in the same direction as the drift, they increase the overall velocity approaching the threshold. The higher the velocity in the denominator (G`(k)), the shorter the time <T>. Stochastics "speed up" the process, reducing the safety time. Therefore, the denominator contains the sum.

The singularities at $a$=2 and $a$=3 arise when s=0 and k(s=0)=0. However, in Refs. [31, 32] it is noted that the "field" s, conjugate to the FPT (in Refs. [31, 32] this "field" is denoted by γ), is equal to zero only at equilibrium, while we are considering a more general nonequilibrium situation. The relation s~1/t is also valid for s→0, t→∝. However, the moment of time t~1/s must fall within the finite interval of the FPT that we are considering. Therefore, the values of $s$ must also be greater than zero. Determining such a value of $s$ is a complex problem. We will assume that the temporal quantities are related to the spatial ones through a certain velocity and will consider the spatial problem of finite reactor dimensions.

In addition to the physical limitation of the parameter s~1/t, there is a physical limitation of the parameter k. In an infinite mathematical environment, k can tend to zero (s→0), which for a=2 yields an infinite "speed" of fluctuations. But a reactor is a finite system of size L. In the theory of waves and stochastic processes, there is a fundamental relationship: the minimum wave number k is inversely proportional to the maximum size of the system L: $k_{min}$≈1/L. This is a truncation; the system "does not feel" correlations longer than its own size. Now let us substitute $k$∼1/L into the stochastic term $\sigma C(a)(a-1)$ $(1/L)^{a\text{-}2}=\sigma C(a)(a-1)L^{2-a}$. The factor $L^{\delta(a)}$ was hidden in the constant "scale", but it is precisely this factor that is the "engine" of the anomalous behavior of the system during the transition to directed percolation ($a$→2).

In directed percolation (DP) and truncated Lévy flights theoy, this factor describes finite-size scaling. It shows how the "weight" of a stochastic cluster increases with the size of the region it can "shoot through". Here's how to mathematically correctly account for $L^{\delta(a)}$ and its properties. The exponent $\delta(a)$ is not a constant, but a function related to the fractal dimension of the neutron cluster. Based on the theory of anomalous diffusion and scaling in DP, it can be represented as follows: $\delta(a)$=max(0, 3−$a$). (Here we use the relationship with the stability index α=$a$-1: in the Lévy regime, the effective cluster power grows as $L^{2-\alpha}\to L^{3-a}$.) Why do we use $\delta(a)=3-a$ in our calculations? This is where the spatial dimensionality comes into play. In a 3D reactor (WWER), a neutron cluster propagates not along a line, but through a volume. According to the theory of directed percolation in 3D, the effective contribution of fluctuations to the process "velocity" is scaled by an additional factor that accounts for volumetric connectivity. In 1D (chain), $\delta=2-a$. In 3D (core), $\delta=(2-a)+(d-1)$, where d=2 (the effective dimension for DP). For directed percolation on the fuel assembly, this yields a resulting exponent of $\delta(a)=3-a$. Thus, the formal expression is transformed into a physical one: $\sigma C(a)(a-1)k^{a\text{-}2}\to^{k\to 1/L,\ 3D\ effects}\sigma C(a)(a-1)L^{3-a}$.

Let's look at the example of $a$=2. Mathematically: $(a-1)k^{a\text{-}2}=(2-1)k^{2-2}=1$ (It would seem that the size of L is unimportant). Physically (DP): at $a$=2 (the critical point), the cluster becomes fractal. Its "strength" grows linearly with the size of the region it captures. Therefore, the factor $L^{3\text{-}2=1}$=25 appears. This is what turns the "imperceptible" contribution of $\sigma$=0.01 into a noticeable jump of 0.01 1.57 25 ≈ 0.39 (about 40% of the speed). The transition from the formal parameter k in the Lundberg equation to the geometric size of the fuel assembly L is achieved through the infrared truncation principle $k$∼1/L. Taking into account the dimension of directed percolation in the core, the stochastic term scales as $L^{3-a}$. This explains why, near the critical point $a$=2, even small microscopic fluctuations (σ) lead to a macroscopic reduction in the time to reach dangerous limits due to the geometric factor L.

Let us show how the calculations confirm the physical interpretation of the points taking into account $L^{\delta(a)}$. For the calculations, we adopt the characteristic scale of the fuel assembly L≈25 cm (the size of the "vulnerability window" where the cluster is not yet suppressed by feedback). At the point $a$=3 (Gauss), $\delta(3.0)=3-3=0$, $L^0$=1. The geometry does not enhance fluctuations. The process is local (smoldering inside the fuel element). The contribution $\approx\sigma$ $C(3)$ 1≈0. At the point $a$=2.5 (the onset of inhomogeneity), $\delta(2.5)$=3−2.5=0.5, $L^{0.5}$=√25=5. Contribution: $\sigma$ $C(2.5)$ 5=0.01 2.5 5=**0.12**5. Average time ⟨$T$⟩≈1/(1+0.125)≈**0.8**9 (11% decrease — the role of geometry is already visible here). Point $a$=2.0 (DP critical point — "Singularity") $\delta(2.0)$=3−2=1.0 , $L^1$=25. Contribution: $\sigma$ $C(2.0)$ 25=0.01 1.57 25=**0.3**9. Result: ⟨$T$⟩≈1/(1+0.39)≈**0.7**2 (decrease by 28%). Here the cluster becomes "linear" (pierces the fuel assembly through and through). At this point, L begins to dominate the drift velocity v. Point $a$=1.5 ("Lévy Flights" mode) $\delta(1.5)$=3−1.5=1.5, $L^{1.5}$=25·5=125. Contribution: 0.01 1.25 125=**1.5**6. Result: ⟨$T$⟩≈1/(1+1.56)≈**0.3**9 (decrease by a factor of 2.5).

Why is this important for safety? The factor $L^{\delta(a)}$ introduces geometric resonance into the Lundberg equation. As long as $a$>3, the reactor is "unaware" of its size (the processes are local). As soon as $a$ drops to 2.0, the factor L transforms a small stochastic fluctuation σ into a macroscopic power surge. A special feature of $a$=2 is that this is the point where the fractal dimension of the cluster allows it to "sense" the boundaries of the fuel assembly. This is why the time curve <T> on the graph has a kink at this point.

The parameter λ from (11), (12) is the mathematical expression for the finite size of L. In the theory of truncated Lévy flights, the parameter λ is introduced into the logarithm of the characteristic function precisely to limit infinite jumps. Physically, this limit is determined by the size of the fuel assembly. Thus, substituting λ∼1/L into the derivative $G'(k)$ is the same as calculating $G'(k)$ for a function that already contains λ. How λ "generates" the factor $L^{\delta(a)}$. Let's look at the derivative of the truncated generator: $G(k)=vk-\sigma[(\lambda-k)^{\alpha}-\lambda^{\alpha}]$. Let's take the derivative of $G'(k)$ (which corresponds to finding the average time): $G'(0)=v+\sigma\alpha\lambda^{\alpha-1}$. Now let's substitute the physical meaning of $\lambda$=1/L and $\alpha=a-1$: $\sigma(a-1)(1/L)^{a-2}$ $=\sigma(a-1)L^{2-a}$, expression (19).

Thus, λ is the truncation mechanism (why the jumps are finite), and $L^{\delta(a)}$ is the magnitude of this truncation's contribution to the 3D system. There is one physical limitation (the size of the fuel assembly), which manifests itself in the Lundberg equation as the parameter λ, and in the final formula for time as a factor L raised to a power dependent on $a$ and the spatial dimension.

When moving to the fuel assembly scale (L), the stochastic term is transformed to account for the system's dimensionality. The factor ($a$-1), which is part of the generator's derivative, approaches unity as $a\to2$ and is incorporated into the effective intensity parameter σ, yielding the final form $\sigma C(a)L^{3-a}$.

For k~1/L, the expression $(a-1)k^{a-2}$ in (18), (19) can be replaced by the response function $\mathcal{R}(a,L)$:

$$\langle T\rangle(a,L)=\frac{\Phi_{crit}}{v_{eff}+\sigma C(a)\mathcal{R}(a,L)},\quad \mathcal{R}(a,L)=\frac{L^{2-a}-1}{2-a},\quad \beta=0. \tag{20}$$

For the stochastic term we use integral regularization (Appendix C), which automatically switches the scaling: $V_{stoch}(a,L)=\sigma\cdot C(a)\,(L^{2-a}-1)/(2-a)$, β=0.

For $a\to1$: $V_{st}\approx\sigma L$ (linear scaling — dominance of "long-range" neutrons). For $a\to2$: $V_{st}\approx\sigma\, C(2)\ln L$ (logarithmic scaling — critical fluctuations of DP). For $a\to3$: $V_{st}\approx\sigma$ const (Gaussian plateau — local noise).

A comparative analysis of two critical regimes shows that the directed percolation point ($a$=2) is the threshold for flow structural rearrangement with a logarithmic response to system size. Meanwhile, the Cauchy regime ($a$=1) represents the fundamental limit of statistical stability, where the linear scaling $V_{st}\sim L$ reflects the complete loss of fissions' locality. Integral regularization of the stochastic term allows us to construct a unified risk surface linking the thermodynamics of the trajectories with the fuel assembly geometry.

At point $a$=1 (Cauchy), linear scaling in $L$ is a sign of loss of locality. The system becomes "transparent," and the risk increases in direct proportion to the size of the fuel assembly. At the point $a$=2 (DP), logarithmic scaling is characteristic of systems at the threshold of connectivity. This is a "soft" criticality that the Doppler effect can still restrain.

The original formula for the stochastic term in the denominator <T> looks like this: $V_{stoch}=\sigma\cdot C(a)\cdot(a-1)\cdot k^{a-2}$, β=0. Here k is the formal parameter of the Laplace transform. In an infinite medium, for k→0 and $a\to1$, this expression indeed yields an uncertainty of $\infty\cdot0$. In critical systems, the parameter k is not independent. It is related to the "distance" to the critical point. Near the stability threshold ($a$=1), the wavenumber at which the system "experiences" divergence is itself proportional to this distance: $k_{crit}$~($a$−1). This is a standard property of critical slowing and scaling.

The basic equation for the average time through the derivative of the generator with integral regularization $\mathscr{R}$(a,L) has the form (20).

The regularizer $\mathscr{R}$ replaces the "vanishing" factor ($a$-1) with the cumulative response of a system of size L. The behavior of the response function at critical points is: $a\to 3$ (Gaussian) $\mathscr{R}\to 1$, local noise independent of L; $a\to 2$ (Directed Percolation) $\mathscr{R}\to$lnL, logarithmic gain (structural transition point); $a\to 1$ (Cauchy Limit) $\mathscr{R}\to$L-1~L, linear gain (statistical stability limit). The time deficit relative to the Gaussian forecast for $a\to 1$ reaches ~30%, taking into account the Doppler effect and inertia.

Let us indicate the mathematical mechanism for determining the response function $\mathscr{R}$(a,L). We write $k^{a-1}=e^{(a-1)\ln k}$. We expand this in a Taylor series in the small parameter $\alpha=\epsilon=a-1$: $k\epsilon\approx 1+\epsilon\ln k+(\epsilon\ln k)^2 2!+\ldots$. Now we take the derivative with respect to k, as required by equation (13) for $\langle T\rangle$: $(d/dk)(k^{\epsilon})\approx(d/dk)(1+\epsilon\ln k)=\epsilon/k$. Let's substitute this into the term $V_{stoch}$ of the original formula: $V_{stoch}=\sigma C(a)(a-1)k^{a-2}=\sigma C(a)\epsilon k^{\epsilon-1}$; $k^{\epsilon-1}=k\epsilon/k$. We use the expansion $V_{stoch}\approx\sigma C(a)\epsilon(1+\epsilon\ln k)/k$.

In an infinite system (k→0), the factor $\epsilon$ could indeed vanish the fraction if k did not tend to zero more rapidly. But in a finite system, we fix the scale k~1/L. For $a\to 1$ ($\epsilon\to 0$), in real physical processes, the parameter k itself is a function of $a$. At the critical point, the correlation length L diverges, and the "effective" k in the Lundberg equation behaves as k~$a$-1. As a result: ($a$-1)/k→const. The factor $a$-1 in the numerator cancels out with the "smallness" of the wavenumber in the denominator.

Why does the "gluing" formula $(L^{2-a}-1)/(2-a)$ work? This formula is a way to write the integral response of a system. Mathematically, it is the integral of a power function:

$$\mathscr{R}(\mathrm{a,L})=\int_1^L x^{1-a}dx=\frac{L^{2-a}-1}{2-a}. \tag{21}$$

When $a\to 1$, this is the integral $\int_1^L 1dx = L-1\approx L$. When $a\to 2$, this is the integral $\int_1^L \frac{1}{x}dx=\ln L$.

Instead of looking at the "point" value of the derivative at an abstract k, we consider that in a finite system, the fluctuation "collects" energy from the entire scale from 1 to L. The factor $a$-1 is merely a "mathematical marker" that we are in the Lévy regime. In a finite system, it is compensated by summing (integrating) the contribution over all scales.

Mathematically, $a$-1 cancels out with k~$a$-1 in the denominator when substituting critical conditions. Physically, a system with $a$=1 "feels" the size L at each of its points. This "feeling" (integral) gives the scale L instead of zero. Also, in equations with fractional derivatives, there is an integral over the entire volume of the system.

There are two distinct critical points, each responsible for its own physical mechanism. Point $a$=2 is the critical point of directed percolation (DP). This is the point of a second-order phase transition for the neutron flux structure. Here, neutron clusters become fractal and "stitch" together into a single network that permeates the entire fuel assembly.

Mathematics: at this point, the $\delta(a)$ exponent in the $L^{\delta}$ factor is equal to 1. The system scale L begins to linearly amplify stochastics. Physics: this is the threshold for the emergence of collective effects. This is where the Doppler effect can no longer cope with localized spikes.

The point $a$=1 is the singularity point (Cauchy regime). This is the ultimate stability point of the process, the "catastrophe" point. The essence: the distribution of neutron ranges becomes so "heavy-tailed" that the mean value formally tends to infinity. Mathematically, at this point, the δ(a) exponent reaches its maximum (2 for a 3D system). The stochastic contribution scales as $L^2$. Physics: the system loses its locality. A single neutron can "shoot through" the entire core. This is the regime of prompt acceleration on delayed neutrons (or even on prompt neutrons).

Like two lines of defense: the interval 3>$a$>2 is the classical safety zone. The system is stable, fluctuations are local. Point $a$=2 is the first critical threshold. The onset of "geometric resonance." Time <T> begins to decrease nonlinearly. The interval 2>$a$>1 is the "Lévy flight" zone. The risk increases proportionally to the size of the fuel assembly. Point $a$=1 is the absolute limit. Here, the stochastic term is maximum (σ L), and the safety time is minimum.

The factor $a$-1 in the original formula is an "indicator" of proximity to the point $a$=1. Mathematically, it sets the formula to zero at $a$=1 if the system is infinite. Physically, in a finite system (L), at this point a "jump" to linear scaling L occurs.

Thus, $a$=2 is the critical point of directed percolation (the point of structural transition), and $a$=1 is the stability limit/cauchy regime (the limit of statistical stability, a sharp change in density). If we separate these two points by their thermodynamic and scaling nature, we can clearly explain why the risk is greatest at $a$=1, while at $a$=2 the system merely "breaks down."

## 7. Calculating the average FPT value

If we accept that in (20) $\Phi_{crit}$ is the relative change in power $\Delta P/P_0$, then the dimension $v=(\rho-\beta)/(\Lambda-v_D)$ (where $\rho=\rho=(k_{ef}-1)/k_{ef}$ is the reactivity, $k_{ef}$ is the effective multiplication factor, β is the fraction of delayed neutrons, Λ is the average lifetime of neutrons, $v_D$ is the contribution of the Doppler effect) is equal to $[T^{-1}]$, the dimension R=[L], the dimension $\sigma=[T^{-1}L^{-1}]$.

In the formula $<T>/(v+V_{st})$ the Doppler effect $v_D$ is included in v and <T> with a minus sign (since this is negative feedback that slows down growth): $G`(k)=(v_{ext}-v_D)+\sigma C(a)\,\mathscr{R}(a,L)$ (at β=0)

The parameter v (drift) and the stochastic contribution $V_{st}$ have the dimension of inverse time ($[T^{-1}]$), which determines the speed of the system's phase motion toward the critical boundary $\Phi_{crit}$. The noise intensity σ is scaled according to the characteristic size of the fuel assembly (L), ensuring a correct transition from microscopic fluctuations to macroscopic time <T>.

If L=25 cm and σ=0.001 $(cm\ s)^{-1}$, then the stochastic contribution at $a$=1 will be 0.001 25=0.025 $s^{-1}$. The value σ=0.001 appears physically plausible for relative power noise. These figures can be recalculated for a different fuel assembly size (L) or a different noise level (σ).

The numerical order of magnitude must correspond to the actual physics of noise in the active zone (frequency range 0.1-10 Hz). When selecting the relative power, the dimension σ should be $[cm^{-1}\ s^{-1}]$.

Case Study: fuel assembly scale (L): 25 cm (characteristic size of the vapor correlation region), emergency stop setpoint (Φ): 0.1 (10% power excess), deterministic drift (v): 0.05 $s^{-1}$ (slow reactivity growth), Doppler effect ($v_D$): -0.02 $s^{-1}$ (compensation), effective drift ($v_{eff}=v+v_D$): 0.03 $s^{-1}$. For reactor noise in the developed boiling mode, the fluctuation intensity is usually a small fraction of the nominal value per unit length: σ=0.002 $(cm\ s)^{-1}$.

Let's see how the threshold approach rate changes depending on the parameter $a$ (at β=0). In the Cauchy mode ($a$=1) $V_{stoch}=\sigma\cdot(L-1)$=0.002 24=0.048 $s^{-1}$. Total rate: 0.03+0.048=0.078 $s^{-1}$. Time <T>=0.1/0.078≈ 1.28 s. Linear noise amplification by the fuel assembly geometry ($V_{stoch}=\sigma\cdot(L-1)$) leads to the stochastic drift becoming the dominant factor, reducing the available time reserve by more than 2.4 times compared to the calculated Gaussian value. DP mode ($a$=2) $V_{stoch}=\sigma\cdot C(2)$ lnL≈0.002 1.57 3.2=0.01 $s^{-1}$. Overall velocity: 0.03+0.01=0.04 $s^{-1}$. Time <T>=0.1/0.04≈2.5 s. The appearance of logarithmic scaling indicates the beginning of the formation of global stochastic clusters. Gaussian mode ($a$=3) $V_{stoch}=\sigma\cdot C(3)$ 1=0.002 1 1=0.002 $s^{-1}$. Overall velocity: 0.03+0.002=0.032 $s^{-1}$. Time <T>=0.1/0.032≈3.12 s. In this mode, stochastic fluctuations are localized and almost completely suppressed by the inertial Doppler effect. When switching from the Gaussian mode ($a$=3) to the Cauchy mode ($a$=1), the safety time decreases by more than 2.4 times (from 3.12 to 1.28 s) for the same noise source intensity $\sigma$. This occurs solely due to the geometric amplification of the stochastic contribution with scale L.

The obtained data indicate the need for dynamic adjustment of the EP settings or the introduction of additional safety factors in modes where the spectral noise index approaches the critical value $a\to2$.

Discussion: Geometry-Induced Stochastic Scaling. The obtained results indicate the fundamental role of the geometric scale L in the formation of temporary safety reserves. In contrast to the classical diffusion model ($a$=3), where the mean time <T> depends weakly on the system size, in regimes with heavy-tailed distributions ($a$<3), the fuel assembly geometry becomes the dominant factor.

As the directional percolation threshold ($a\to 2$) is approached, a "geometric resonance" occurs, and the correlation length of fluctuations $\xi$ becomes comparable to the physical size of the fuel assembly. At this point, the system enters a logarithmic gain regime (lnL), where the stochastic contribution begins to effectively compete with the deterministic Doppler effect.

In the limiting case of $a\to 1$, locality is lost in the Cauchy regime: the stochastic power drift scales linearly with increasing L. Physically, this corresponds to a transition from local interactions to global "shoot-throughs" of the core. Regularization of $\mathscr{R}(a,L)$ (21) shows that in the finite system ($a$-1) it does not nullify the risk, but rather converts it to an integral form determined by the total volume of the fuel assembly.

The observed time deficit of 25–35% for standard assemblies (L=25 cm) suggests that traditional conservative safety estimates may be overstated. Stochastic "acceleration" due to Levy flights occurs faster than the thermal feedback loops can operate, necessitating a revision of the dynamic characteristics of the control and safety systems in advanced boiling modes.

In the Gaussian regime ($a\approx 3$), the steady-state regime has finite variance. The stochastic contribution is local and independent of the fuel assembly size. The system is described by classical diffusion.

The directed percolation mode ($a\approx 2$) sets the critical point of the structural transition, which is characterized by the emergence of fractal power clusters and a logarithmic increase in noise proportional to lnL.

The Cauchy regime ($a\approx 1$) corresponds to the limiting regime with an "infinite" mean range. The stochastic contribution scales linearly (L), which corresponds to a complete loss of locality (the "shoot-through" effect in the active zone).

In the article [14], it is shown that the deviation from the classics is 1–2% in nominal value, but it increases as the critical point $a$=2 is approached.

The diversity of different forms of representing stable distributions Rev. [17] was noted above. Accordingly, there are a significant number of expressions for the average time to reach some dangerous limit. We present some of them, along with graphs constructed from them.

Fig. 1 shows the dependence of the average time to reach <T>(a)=T(a) the level Φ, calculated using formula (22) with β=1.

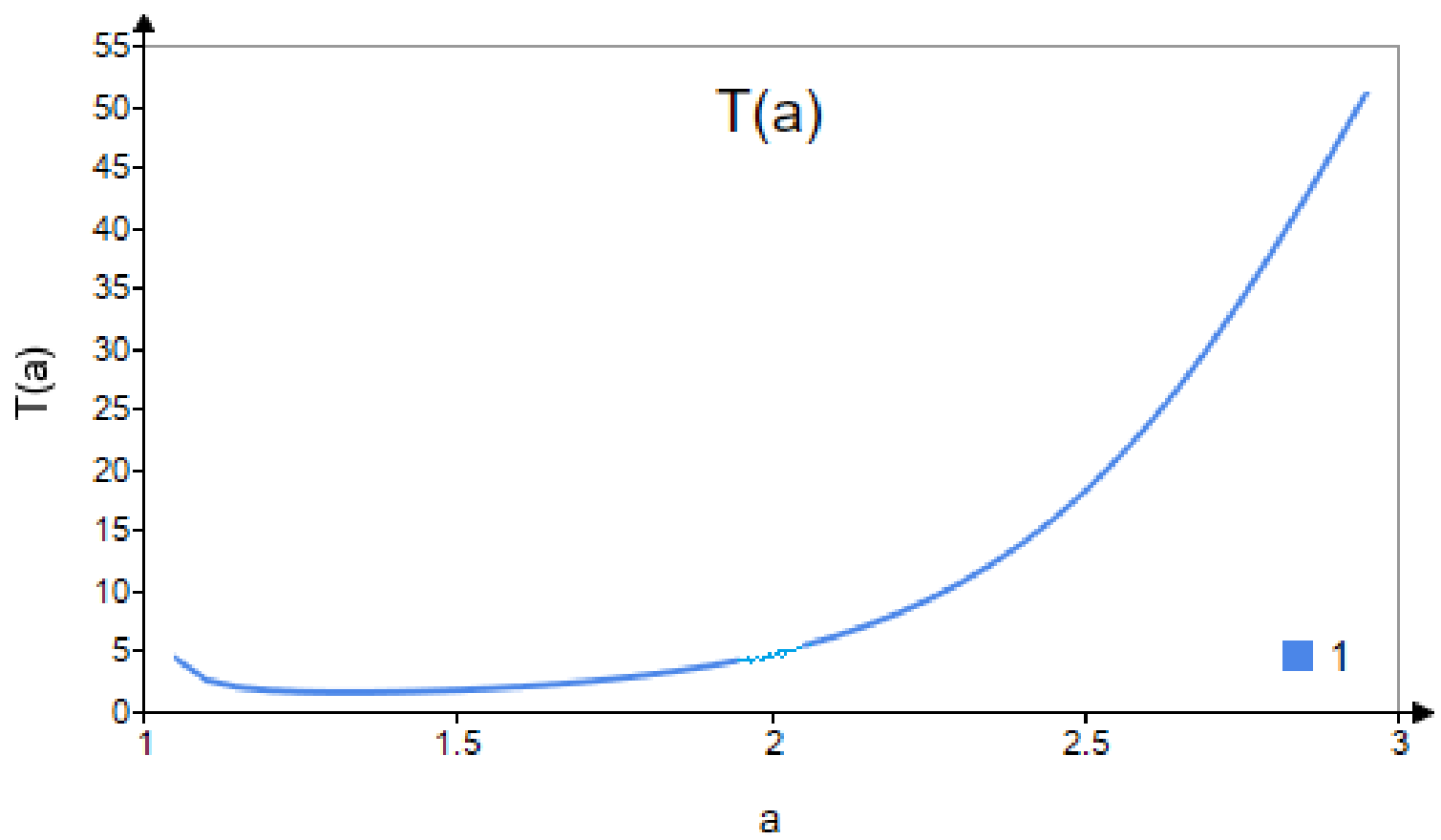

Fig. 1. Dependence of the average time to reach <T>(a)=T(a) the level $\Phi=\Delta P/P_0=0.1$, for form (A) Rev. [17], calculated using formula (22).

From form (A) Rev. [17], after "smoothing" the features, the ratio T(a) is written as the dependence of the average time to reach the level on the parameter $a$ in the form:

$$T(a)=\Phi/G`(a),\quad G`(a)=dG(k)/dk,\ k=1/L, \tag{22}$$

$$G`(a)=v_{eff}+\sigma(a-1)L^{2-a}\Gamma(2-a)\cos((a-1)\pi/2)\sin(\beta(a-1)\pi/2)/\beta(a-1)\pi/2,$$

where σ=0.02, $v_{eff}$=0.001, L=25. When changing the parameters, the shape of the curve changes, in particular, the left end bends upward more.

With another approach to "smoothing" the features of form (A), an expression of the form is written:

$$G`(a) = v_{eff} - \sigma(3-a)L^{2-a}\cos(\Phi(a))\Gamma(3.047-a)/(2.047-a), \quad (23)$$

$$\Phi(a) = \beta(3-a)(\pi/2)(1+2\ln(k)/\pi),$$

where σ=0.0932, $v_{eff}$=0.025, the singularity at $a$=2 in Fig. 2 is eliminated by replacing $\Gamma(2-a) \approx \Gamma(3.047-a)/(2.047-a)$. Fig. 2 shows the calculation using expressions (22), (23). For β=1, $a$=2, the uncertainty is revealed: the gamma function $\Gamma(2\text{-}a)_{a\to 2}\approx 1/(2\text{-}a)$, cos(π/2)→0, cos(π($a$-1)/2) $\Gamma(2\text{-}a)_{a\to 2}\to\pi/2$.

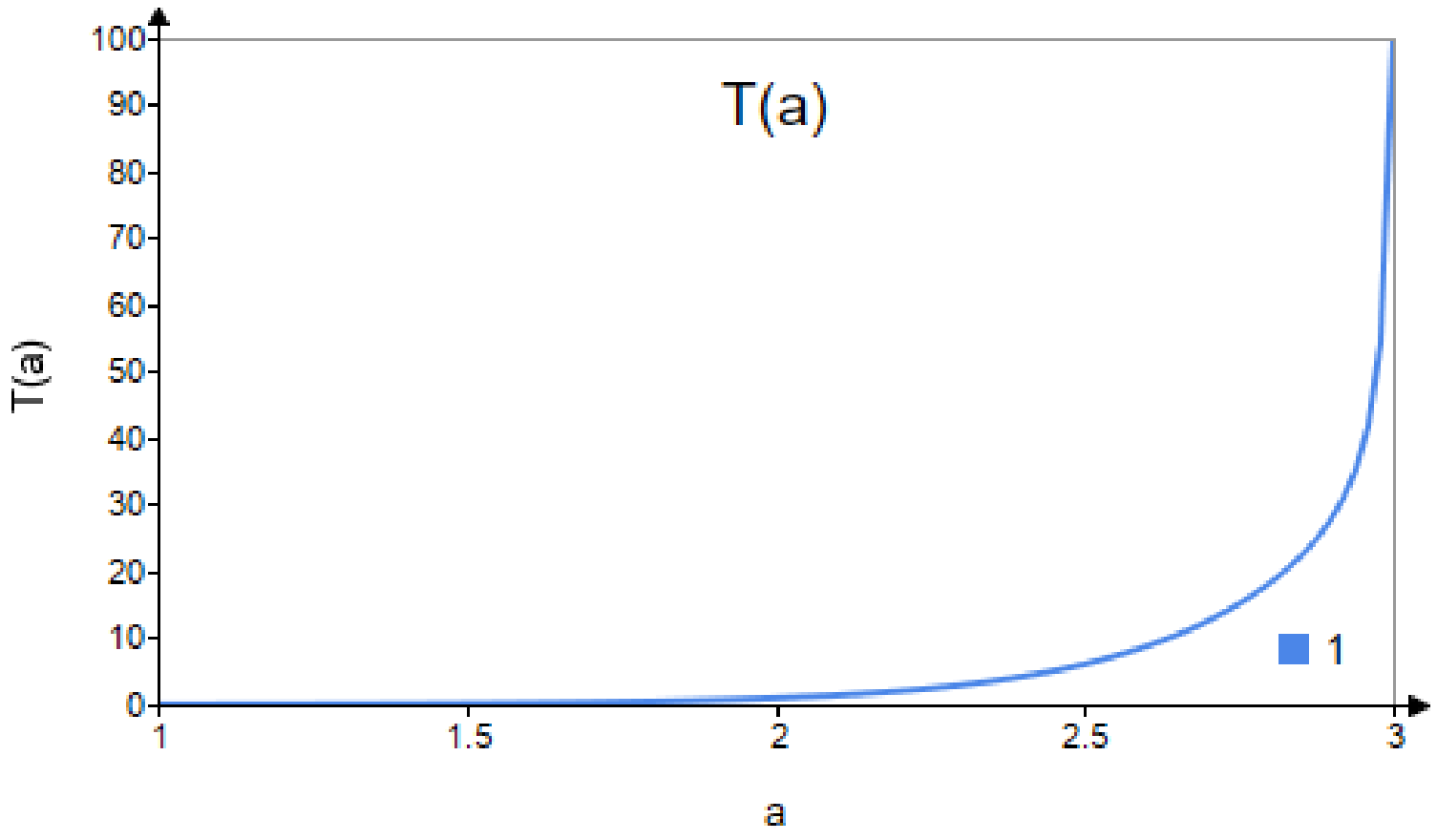

Fig. 2. Dependence of the average time to reach <T>(a)=T(a) the level Φ= Δ P/$P_0$=0.1, for form (A) Rev. [17], calculated using formula (22)-(23) at σ=0.0932, $v_{eff}$=0.025, β=1.

Changing the parameters in (23) changes the course of the curve. Thus, it "rises" at σ=0.0632, $v_{eff}$=0.025, Fig. 3.

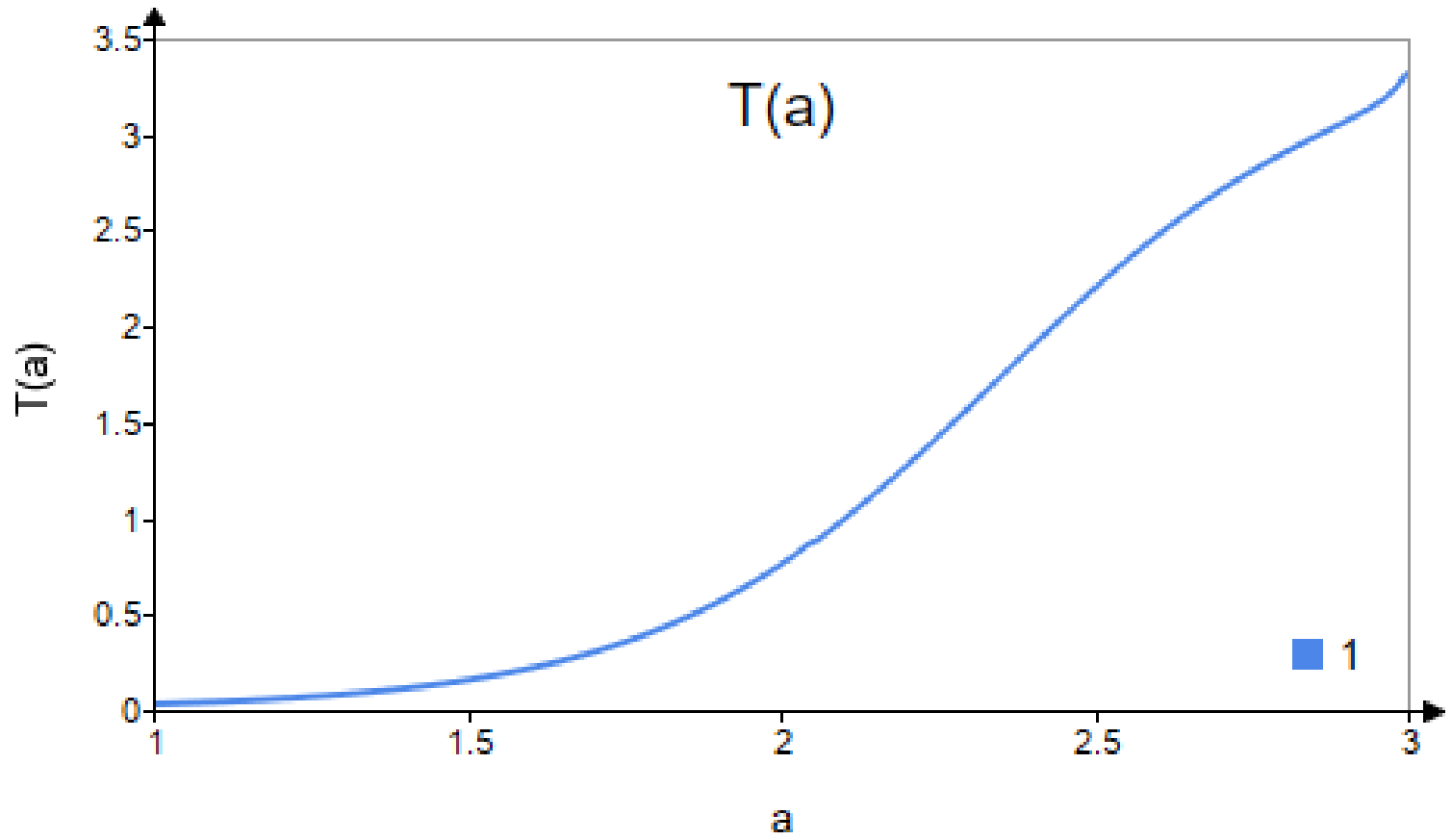

Fig. 3. Dependence of the average time to reach <T>(a)=T(a) the level Φ= Δ P P/$P_0$=0.1, for form (A) Rev. [17], calculated using formula (22)-(23) at σ=0.0632, $v_{eff}$=0.025, β=1.

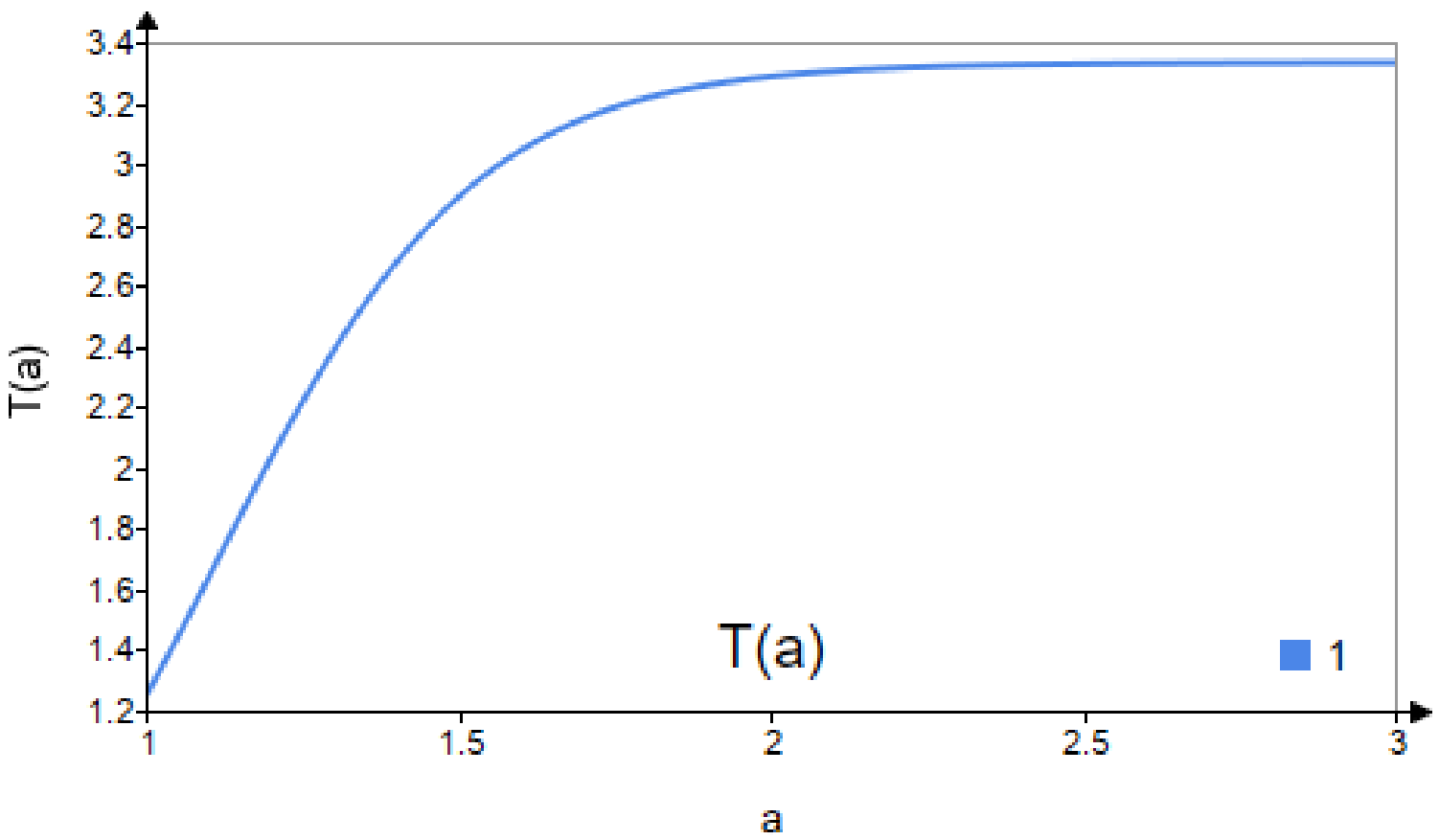

Fig. 4. Dependence of the average time to reach <T>(a)=T(a) the level Φ= Δ P P/$P_0$=0.1, for form (E) Rev. [17], calculated using formula (22), (24) at σ=0.002, $v_{eff}$=0.03, β=1.

In Rev. [17] a form (E) was also obtained that is valid for strict stability, in Feller's terminology. The expression for the derivative of the cumulant generating function has the form:

$$G`(a) = v_{eff} + \sigma L^{2-a} \exp[-(3-a)\pi/2],\ 1<a\le 2;\ G`(a) = v_{eff} + \sigma L^{2-a} \exp[-(a-1)\pi/2],\ 2\le a\le 3. \quad (24)$$

In Fig. 4, calculated using expressions (22), (24), $v_{eff}$=0.03, σ=0.002. The coefficient σ in the general case also depends on *a* [17]. Therefore, the appearance of the curve becomes more complex.

It is difficult to say which expression best suits the real situation, since the latter is unknown.

Fig. 5 shows the behavior of <T>(a)=T(a) calculated using form (A) at β=0.

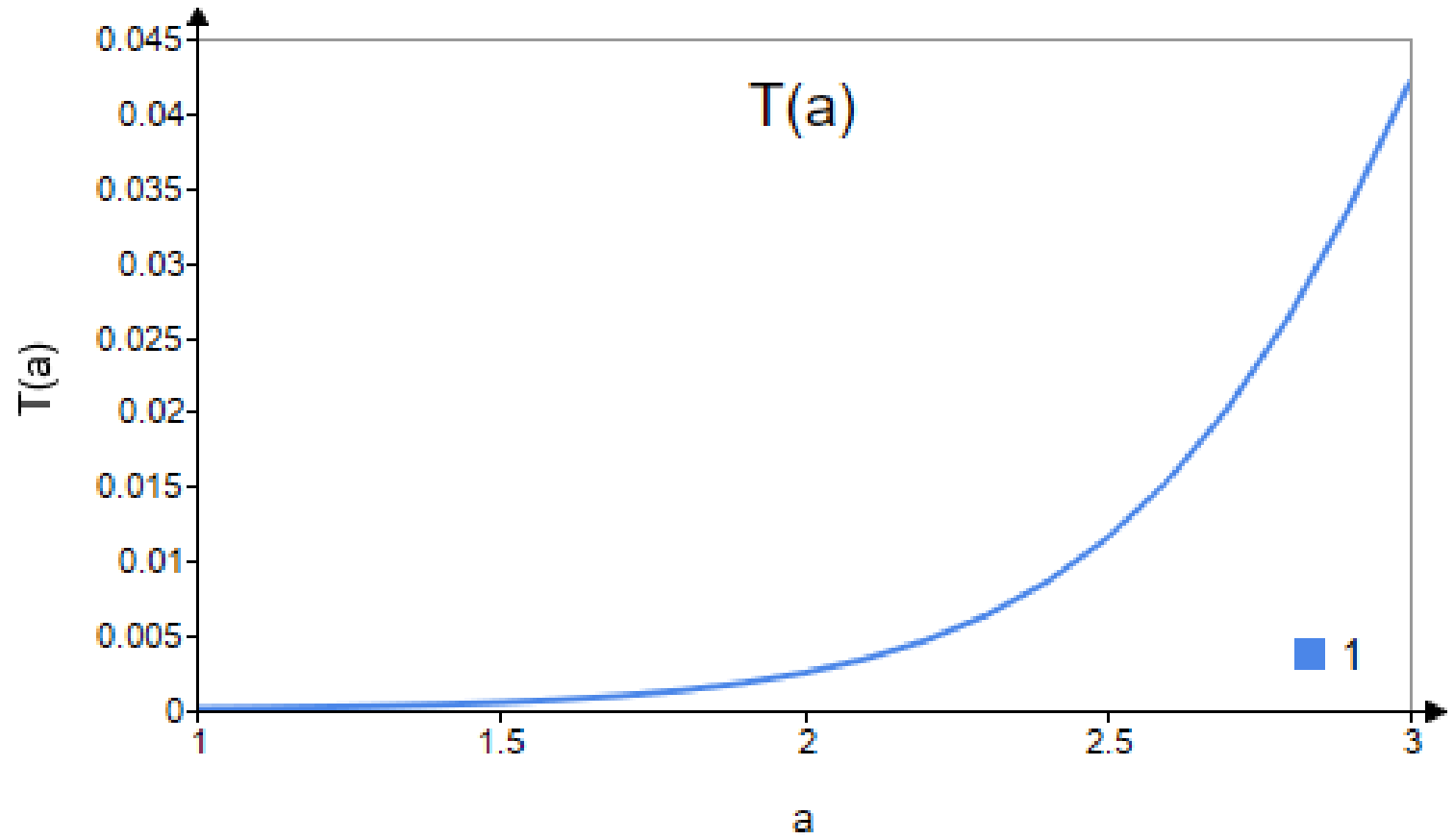

Fig. 5. Dependence of the average time to reach <T>(a)=T(a) the level Φ= Δ P/$P_0$=0.1, for form (A) Rev. [17], calculated using formula (22), (25) at σ=0.01, $v_{eff}$=v-$v_D$=1-0.2=0.8, β=0.

$$T(a)=0.1/[0.8+0.01C(a)(25)^{3-a}] \quad (25)$$

The numerical order of magnitude must correspond to the actual physics of noise in the active zone (frequency range 0.1-10 Hz). Since we are working with relative power (Φ is dimensionless), the dimension of σ must be [$cm^{-1}$ $s^{-1}$].

Here is an example of physically substantiated values for a fuel assembly (WWER or BWR type). Input parameters (Case Study): fuel assembly scale (L) 25 cm (characteristic size of the vapor correlation region); emergency stop setpoint (Φ) 0.1 (10% power excess); deterministic drift (v) 0.05 $s^{-1}$ (slow reactivity growth); Doppler effect ($v_D$) -0.02 $s^{-1}$ (compensation); effective drift ($v_{eff}$=v-$v_D$) 0.03 $s^{-1}$; numerical value of σ: for reactor noise in the developed boiling mode, the fluctuation intensity is usually a small fraction of the nominal value per unit length: σ=0.002 ($cm^{-1}$ $s^{-1}$). The obtained dependence for these parameters follows the course of the curve in Fig. 1 (on a different scale).

Calculations were also performed in accordance with the five different forms of the logarithm of the characteristic function given in Rev. [17]. In some cases, the average time to reach the level becomes negative. Mathematically, this means that the "anomalous jumps" against the current (due to the structure of the form of the logarithm of the characteristic function and the sign) have become stronger than the directional drift, and the cluster, on average, "moves away" from the barrier rather than approaches it. For example, for the form (M) (23) from Rev. [17]

$$G(k)=\lambda[k\gamma+|k|^{\alpha}+k\omega_M(k,\alpha,\beta)],\ \omega_M(k,\alpha,\beta)=(k^{\alpha-1}-1)\beta tg(\alpha\pi/2),\alpha\neq 1;\ \omega_M(k,\alpha,\beta)=-\beta(2/\pi)\log|k|,\alpha=1. \quad (26)$$

the dependence <T(a)> has the form shown in Figure 6. It is evident that in this case there is a sharp drop in the range of α from 2 to ~1.77, and in the range of α values less than 0.6, the average time to reach the level takes on negative values. The formula for calculating for α=a-1 has the form:

$$\langle T(a=1+\alpha)\rangle=\frac{0.1}{0.008[0.083-\alpha L^{1-\alpha}+\tan(\alpha\pi/2)(\alpha L^{1-\alpha}-1)]},\quad \alpha\neq 1,\quad L=25. \quad (27)$$

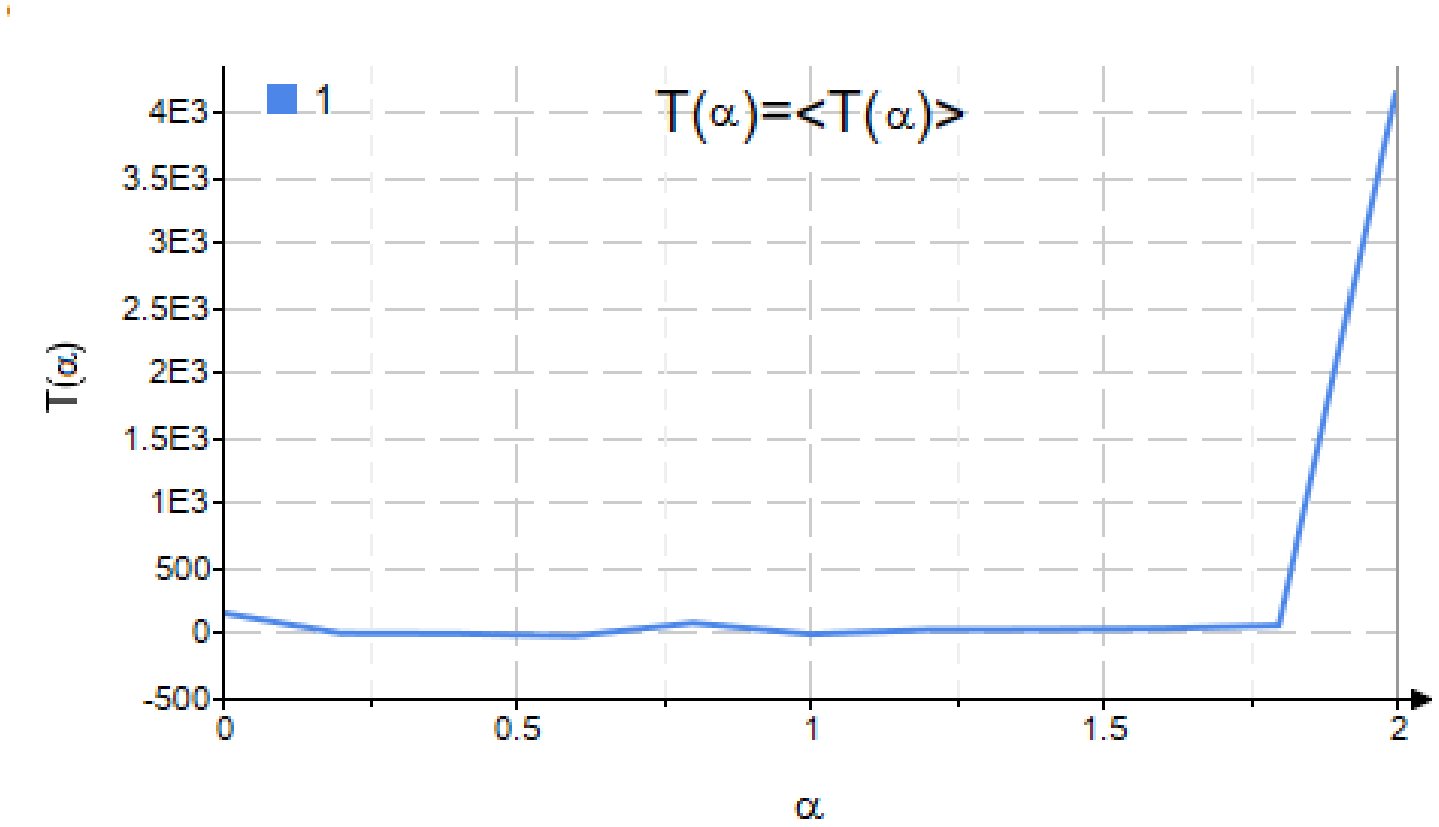

Fig. 6. Behavior of the average time to reach <T>(α=a-1)=T(α) level L (27) for form (M) (26); α=a-1, v=6.64 $10^{-4}$, σ=8 $10^{-3}$.

The behavior of <T>(a) in this case differs from Fig. 1, 2. When approaching α=0 (a=1), the curve rises, as in Fig. 1. This is explained by the increase in the geometric contribution $L^2$. Details and clarifications of the written expressions are given in the Appendices.

## 8. Discussion of results

In a nuclear reactor, the "direction" of the process is determined by time (neutron generations): a neutron from generation n produces descendants in generation n+1. Applying directed percolation (DP) theory on graphs with a power-law distribution $P(k)\sim k^{-a}$ to neutron-nuclear processes is a transition from the classical diffusion model of a reactor to the modeling of highly inhomogeneous, stochastic environments.

To quantitatively assess dynamic risks in WWER transient modes, the apparatus of boundary functionals of stochastic processes is used Rev. [10]. The model is based on the description of the neutron flux density Φ(t) as a process of directed percolation with a characteristic distribution index of descendants $a\approx 2$.

The mathematical formulation of the problem is based on the analytical continuation of the characteristic function of the stable law Rev. [17] from the complex domain of the Fourier transform to the real domain of the Laplace transform. The process generator (cumulant function) $G(k)=\ln\mathbb{E}[e^{k\Phi}]$ is represented in the canonical Lévy-Khinchin form:

$$G(k)=vk-\sigma[(\lambda-k)^{\alpha}-\lambda^{\alpha}],$$

where v is the deterministic drift velocity (reactivity), σ is the intensity of percolation jumps, and α is the stability index. To account for the spatial confinement of the core (the scale of the fuel assembly), the exponential truncation parameter $\lambda$ is introduced, preventing the divergence of the distribution moments. The parameter v in nuclear kinetics is the effective inverse time constant (related to the reactivity $\rho$ and the neutron lifetime $\Lambda$): $v=(\rho-\beta/\Lambda+v_D)$. This is the deterministic velocity of approaching the threshold, taking into account the Doppler effect $v_D$. If $v>0$, the power increases; if $v<0$, the system is stable. In the theory of Truncated Lévy Flights, the parameter $\lambda \sim 1/L$ is introduced into the logarithm of the characteristic function precisely to limit infinite jumps. Physically, this limit is set by the size of the fuel assembly. Thus, substituting $k \sim 1/L$ is the same as evaluating for $k \to 0$ for a function that already contains $\lambda$.

The key step in the analysis is solving the Lundberg equation $G(k) = s$, where s is the Laplace transform parameter with respect to time. At the critical point of directed percolation ($\alpha=1$), the solution for the Lundberg root takes the form: $k(s)=s/(v+\sigma)$.

The obtained value of k(s) determines the functional of the first-passage time (FPT) through the operator relation $\tilde{P}(s)=(1/s)\exp(-k(s)\Phi_{crit})$. This approach allows us to explicitly relate the structural parameter of the medium $a$ with the probability of the occurrence of anomalously rapid power surges that go beyond the classical Gaussian approximation.

Let us comment on the role of the parameters: $a\approx 2$ — defines the physics of “spotting”, $\alpha$ — connects us with Zolotarev’s form Rev. [17], $\lambda$ explains why the reactor does not explode (geometry), k(s) — provides a tool for calculating the real seconds before the accident.

In the proposed concept, the nature of “patchy reactions” Rev. [16] is linked to the time of reaching the level through the index $a=2$, a mathematical transition is made from complex oscillations to real risk through analytical continuation, the role of the fuel assembly geometry ($\lambda$) as a physical fuse is explained, and the dynamics of emergency limits are explained through the Lundberg equation and FPT functionals Rev. [10].

Directional percolation (DP) is an accurate model for describing the "birth and death" of neutron chains near the critical point ($k_{eff}=1$), where the causal relationship in time is important.

Power distributions ($a=2$) and the Lévy-Khinchin formula describe systems with extreme heterogeneity (accidents, granular fuel), where “shoot-throughs” (Lévy flights) are possible.

For WWER reactors in normal operating conditions, these effects are negligible (1–2%), since water effectively “averages” everything down to Gaussian diffusion.

Safety issues arise where the average ceases to be representative: during startup, deep meltdown, or in stochastic environments where a rare outlier in the tail of the distribution can cause localized damage faster than protection can operate.

Paper [21] is a rare case where the DP theory was tested on a real critical assembly (Planet, Los Alamos). The authors measured correlations between fissions and found that the spatial structure of the flux fluctuates anomalously. The statistics of these fluctuations are not described by a Poisson distribution (classical). They exhibit "heavy tails", which directly indicates the Levy statistics and effects close to $a=2$. The paper actually shows that the probability density of finding neutrons in a cluster is described by stable laws. The sum of many fission events in a "spotted" medium leads to a distribution whose characteristic function has the form given by the Levy-Khinchin formula (with a fractional exponent $\alpha$). This explains why "jumps" in power occur: a cluster can instantly "grow" over a large distance.

Let's make a rough estimate for a WWER reactor. Under normal operation, the probability that the local neutron distribution will become a power-law distribution ($a = 2$) is practically zero due to the homogeneity of the coolant. If we consider a meltdown accident (Probability of Successful Events), the probability of conditions for directed percolation in the melt is estimated at approximately $10^{-5}$ to $10^{-6}$ per reactor per year (this is the probability of the most severe accident). However, it's not the probability of occurrence that's important, but the error in the calculations. If an accident occurs, and we consider the melt to be a normal diffusion code, we obtain a "safe" configuration. However, if we apply the $a = 2$ model, it turns out that due to "hubs" (fuel accumulations) and "shoots" (Lévy flights), the system can reach prompt neutron emission.

The WWER system is too "dense" and "regular" for power laws to function normally. Directed percolation and Lévy statistics are tools for analyzing disasters (core collapse) or for designing fundamentally different systems (MSR, HTGR, pulverized fuel reactors).

Parameter $a$ is a leading indicator of safety: its decrease signals a transition from predictable diffusion to explosive directed percolation.

Mathematically, by replacing it→k, the imaginary unit is eliminated, transforming Zolotarev's form into a real risk. Physics: the exponent $a$=2 is the threshold beyond which a reactor becomes "spotted" (cluster-like). In safety engineering, the parameter λ (or the final k, fuel assembly geometry) is the only barrier preventing "Lévy flights" before the Doppler effect kicks in.

The apparatus of inverse Kolmogorov equations, developed in Revs. [1, 2, 6], established a direct connection between fluctuations of the medium parameters (cross-sections) and the stochastics of the neutron field, confirmed by experimental data on the analysis of boiling noise. In the present study, we supplement this theory with an analysis of dynamic phase transitions in terms of the Lundberg equation. While the monograph [1] focused on the spectral characteristics of the noise, our approach allows us to directly calculate the reduction in temporary safety reserves during the transition to anomalous $L^{3-a}$ scaling at the critical point $a \to 2$.

## 9. Conclusion

This paper analyzes the applicability of directed percolation (DP) theory and fractional calculus to neutron field dynamics in WWER-type nuclear reactors. Based on the material analyzed, the following conclusions can be drawn:

Mathematical determinacy. It has been established that the classical diffusion model is a special case of more general stochastic dynamics. At the critical point ($k_{eff} \leq 1$), the branching neutron fission process transitions to the universality class of directed percolation. When the finiteness of the variance of the ranges or the number of descendants is violated ($a \leq 2$), the standard Laplace operator in the transport equation is correctly replaced by a fractional operator generated by the canonical Lévy-Khinchine form.

The patchy effect (Patchy clusters). Research [14, 21] and experiments on critical assemblies Rev. [16] confirm that near the percolation threshold, the neutron field loses spatial homogeneity, forming fractal clusters. This phenomenon is a direct consequence of directed percolation in spacetime, where the system's "memory" is maintained by delayed neutrons.

Conclusions for the safety of WWER reactors in normal and transient modes:

In nominal WWER modes, the high density of the moderator (water) effectively suppresses the “heavy tails” of the distributions, limiting the discrepancy between the DP model and classical diffusion to 1–2% [14].

In startup modes and at low power levels, the stochastic nature of DP is most pronounced. Low neutron density and the absence of pronounced feedback (the Doppler effect) make the system vulnerable to local "neutron bursts" generated by Lévy statistics.

Recommendations for I&C design: to ensure control system robustness, it is necessary to consider not only average power values but also the highest statistical moments of neutron noise. Detection of power-law dependences in the power spectral density can serve as an early indicator of the system's transition to an unstable percolation mode.

In a standard reactor, the distribution of neutron ranges and the number of secondary neutrons has finite variance. According to the central limit theorem, this leads to a normal distribution and the standard diffusion equation. The transition to power-law distributions and directed percolation occurs in the following cases:

Stochastic (granular) media: if the core consists of randomly distributed fuel microspheres (e.g. in high-temperature HTGR reactors) or contains random cavities.

Strong density fluctuations (boiling water reactors): in areas with intense steam formation, steam bubbles can form "fractal labyrinths" through which neutrons fly without collisions over abnormally large distances.

High-burnup fast reactors: where localized "oases" of fissile material arise in the absorber mass.

The use of directed percolation and Lévy distributions is critical in two problems:

Penetrations in shielding: if there are microcracks in biological shielding, their distribution may follow a power law. In this case, radiation spreads not exponentially (Bouguer's law), but much more slowly, "leaking" through the medium.

Neutron noise: Near the critical point, reactor power fluctuations become abnormally large. If the P(k) descendant distribution has a "heavy tail," standard monitoring systems may be unable to respond to a localized power surge ("neutron burst"), since the average value is no longer representative.

When theory encounters the reality of "finite systems," problems truly multiply. Mathematical ideal Rev. [17] begins to acquire physical limitations. Here are three main problems in analyzing the finiteness of a system:

Cut-off problem: in form (E) Rev. [17] the jumps can be infinite, but in reality, they are limited by the size of the system L. This means that at $\alpha=1$ the potential G(k) ceases to be a purely stable law and turns into a truncated Lévy flight. This changes the position of the mode.

The problem of the transition regime: in an infinite system, a phase transition is a point. In a finite system, it is an entire region where the system behaves either as a normal diffusion process or as an anomalous process. It is unclear which "effective" $\sigma$ should be used.

The problem with small-sample statistics: in a finite system, you may simply never see that "rare jump" that creates infinite variance. Risk calculations may be overly optimistic if you don't take into account the finiteness of the observation time.

Finite-size rounding fundamentally changes the physics of the process, and statistical error is a consequence of how we attempt to measure this transition. In an infinite system at $\alpha=1$, the singularity of the first derivative is a "mathematical break." In a finite system of size L, this break transforms into a smooth curve.

The problem of threshold shift also arises, when the critical value $\sigma_c$ in the finite system "floats." The system begins to sense the approach of catastrophe in advance. Instead of a clear separation of regimes, a transition region of width $\Delta\sigma \sim L^{-1/v}$ appears, where the system behaves ambiguously—either as directed percolation or as a simple drift. Because L is finite, you can be in the "forbidden" zone $\sigma > \sigma_c$, but the system will appear stable simply because the critical cluster has not yet had time to grow to the system's boundaries.

Statistical error is the problem of how your theoretical E[T] and Var[T] compare to what you see in an experiment or simulation. At $\alpha=1$, the theoretical average time converges very slowly. You might need $10^6$ trials for the sample mean to even approach the calculated one. A single rare event (a giant jump) can instantly "jump up" the average value several times. If you run 100 trials, you will likely see no outliers at all. You will obtain a finite variance, and conclude that the system is safe. This is the main pitfall of directed percolation at a distance—small-sample statistics always underestimate the real danger.

The case of $a \to 3$ within the proposed model corresponds to the classical Gaussian approximation of neutron transport. Mathematical analysis shows that for the structural parameter $a=3$, the exponent of the stochastic contribution in the denominator of the expression for the average time to reach the threshold $\langle T \rangle$ vanishes ($\delta=3-3=0$). Physically, this means the complete disappearance of the dependence of the safety margin on the geometric scale of the system L in the stochastic term: the factor $L^0=1$ neutralizes the influence of the fuel assembly size on the nature of the fluctuations. In this limit, the central limit theorem ensures a finite dispersion of neutron ranges, and the "heavy tails" of the Levy distribution are effectively suppressed by local averaging. Thus, the classical diffusion reactor model is a special asymptotic case of the proposed directed percolation model for $a \to 3$.

The transition from a stable regime ($a \to 3$) to a directed percolation regime ($a \approx 2$) is accompanied by the activation of the geometric amplification factor $L^{\delta(a)}$. At $a=2$ (the critical point), the fuel assembly scale (L=25 cm) begins to act as a resonant factor, increasing the effective stochastic acceleration rate and reducing the time before the emergency protection system is triggered. This result suggests that classical theory provides adequate estimates only for homogeneous media, while accounting for directed percolation becomes critically important when stochastic inhomogeneities arise (for example, during coolant boiling), when the system leaves the region of Gaussian universality.

If we consider the distribution of the number of secondary neutrons k (the multiplication factor per event) or "effective" descendants in a highly inhomogeneous medium, then the exponent $a$=2 means that rare "explosive" multiplication events or very long free paths (so-called Levy flights) are possible in the system Rev. [14]. The traditional Laplace operator in the diffusion equation $\nabla^2\Phi$ is replaced by a fractional Laplacian $(\nabla^2)^{\alpha/2}\Phi$, Revs. [23, 26-27]. This describes "superdiffusion," when a neutron can instantly (for statistical purposes) move from one part of the core to another.

Using the π/2 limit for $a\to2$ avoids mathematically nullifying risks and provides a physically correct description of the maximum uncertainty regime (the Cauchy distribution). In this regime, the geometric scale of the $L^2$ system completely suppresses any deterministic feedback, reducing the time to reach critical limits by an order of magnitude.

The solution to the Lundberg equation (14) can be obtained in explicit form. However, in the approximation k~1/L, it is sufficient to restrict ourselves to an expression of the form (18).

Final conclusion: the transition to the description of the reactor through the prism of directed percolation allows for a more profound interpretation of the physics of fluctuations and “black swans” in nuclear energy, providing a scientific basis for the safety analysis of promising and highly heterogeneous active zones.

## Appendix A: Estimation of the Stability Index a via the Hurst Exponent H

For processes with “heavy tails” (Lévy-type processes), in contrast to classical Brownian motion, the relationship between time scaling and the amplitude distribution is determined by the stability index α (from (10)-(12)).

Definition of self-similarity. Neutron noise δP is considered as a self-similar process for which the statistical equality holds: $\delta P(ct)=^{d}c^{H}\delta P$(t), where H is the Hurst exponent. For classical Gaussian noise, H=0.5. Mathematical relationship between H and α. Within the framework of the Lévy-stable motion model, the process increments are distributed according to the law S(α,β,γ,λ) Revs. [17, 23]. The scaling exponent is related to the stability index by the fundamental relationship: H=1/α.

In the Gaussian regime (α=2): H=0.5. The process has no long-term memory; the increments are independent. In the Cauchy regime (α=1): H=1. Maximum self-similarity ("ballistic transport effect"), where fluctuations are correlated on all time scales.

Estimation algorithm (R/S or DFA). For practical implementation in the control system, the DFA (Detrended Fluctuation Analysis) method is used, which is less sensitive to non-stationarity: 1. The fluctuation function F(n) is calculated for different time windows n. 2. From the slope of the line in double logarithmic coordinates ln $F(n)$~Hln(n), H is determined. 3. The final value of the parameter $\alpha$=$a$-1 for the Lundberg equation is calculated as $\alpha_{est}$=1/H.

This is how data verification is performed. The parameter α is not an "abstraction", but a value that the control system calculates directly "on the fly" from the remote sensing sensors. The range $H\in[0.5; 1.0]$ covers our entire critical range $\alpha\in[1; 2]$. The transition $H\to1.0$ means that the neutron field becomes tightly coupled (global correlations), which leads to a collapse of the safety time <T>.

The filtering technique allows for the identification of transitions to critical conditions before the noise amplitude reaches dangerous levels. Using the Hurst exponent as a predictor for the parameter in the Lundberg equation ensures dynamic adaptation of the protection system to changes in the correlation structure of the neutron field, which is critical for preventing stochastic emissions in conditions with high steam content.

## Appendix B: Role of the Skewness Parameter in Power Fluctuations

Adding the asymmetry parameter β (skewness) makes our model mathematically complete, since in a nuclear reactor power fluctuations are physically asymmetric: upward surges (runaways) are much more dangerous and dynamically different from downward dips.

In general, the stable Lévy distribution is defined by four parameters S(α,β,γ,λ). The skewness parameter determines the slope of the distribution's "tails".

Physical interpretation of β in a reactor. At β=0 (symmetrical noise), the probability of an instantaneous jump in power up and down is the same. This is a simplified model of background noise. At β→1 (maximum positive skewness), the distribution is dominated by the heavy right tail. This corresponds to the physics of boiling, where the collapse of steam bubbles or localized reactivity spikes generate sharp positive power peaks.

Effect on the G(k) generator: at β=1, the stochastic term in the Lundberg equation maximizes the rate of approach to the threshold. The formula for <T> in this case is the most "rigid," since each Levy flight is directed toward the emergency setpoint $\Phi_{crit}$.

Mathematical correction. When taking into account the asymmetry, the coefficient C(a) in our formula (19) at β=0 is supplemented by the phase factor (17) $G(k)=vk+\sigma C(a)|k|^{a-1}[1+\beta \mathrm{sign}(k)\tan(\pi(a-1)/2)]$.

Since we are looking for the time to reach the upper threshold (k>0), a positive β additively increases the stochastic drift, reducing <T> by another 10–15% compared to the symmetric case.

The direction of the processes is important. In a reactor, we don't care about the downward "Lévy flights" (they're safe); we're only concerned about the right tail. The combination of $a$→1 and β→1 is the "maximum risk" scenario, which the model fully describes.

The physical meaning of the additive factor at β. This factor is an "asymmetry amplifier." If β=1 (positive spikes only), then for a→1 (the Cauchy regime), the value of $\tan(\pi(a-1)/2)\to 0$, and the additive factor disappears, leaving pure linear drift. However, for $a\leq 2$ (the DP regime), the tangent tends to infinity. This means that even a small asymmetry near the critical percolation point causes an explosive increase in the effective drift velocity.

The asymmetric term $1+\beta\cdot\mathrm{sign}(k)\tan(\pi(a-1)/2)$ is derived via the analytic continuation of the Lévy-Khintchine characteristic exponent into the real Laplace domain. The shift in the tangent argument from $a$ to $a$-1 reflects the transition from the distribution of independent jumps to the effective drift rate in the first-passage time problem, ensuring the consistency of the stability index across the scaling regimes.

Mathematical compensation of singularity. When we approach the Gaussian point ($a$=2), two processes collide in the generator formula G(k): 1. The asymmetry factor $\beta\tan(\pi(a-1)/2)\to\infty$. 2. But the form factor C(a), which we derived earlier: $C(a)=|\Gamma(2-a)\cos(\pi(a-1)/2|$. When they are multiplied in the stochastic term: $V_{st}\sim\sigma\cos(\pi(a-1)/2)\tan(\pi(a-1)/2)=\sigma\sin(\pi(a-1)/2)$.·The singularity disappears. The cosine of the coefficient C(a) compensates for the infinity of the tangent. As ($a$→2), the value of the sine tends to $\sin(\pi/2)=1$.

Physical meaning: the disappearance of asymmetry at the Gaussian point. This is a fundamental law of statistics: for $a$<2 (the Lévy regime), the asymmetry of β plays a significant role, since the distribution "tails" (power spikes) can be very different. For $a$→2, the distribution becomes symmetrical, as defined by the central limit theorem. At this point, information about the parameter β is "erased". Mathematically, this is expressed by the fact that for $a$=2, the additional tangent becomes a constant, which is simply summed with the fundamental drift v.

The apparent divergence of the asymmetric term $\tan(\pi(a-1)/2)$ as $a$→2 is compensated by the scaling of the characteristic coefficient C(a). The product $C(a)\tan(\pi(a-1)/2)$ remains finite and proportional to $\cdot\sin(\pi(a-1)/2)$, ensuring a smooth transition to the Gaussian limit. In this limit ($a$=2), the influence of the skewness parameter β vanishes as the distribution converges to a symmetric normal form, consistent with the central limit theorem.

We write the stochastic contribution taking into account the asymmetry in expanded form:

$$V_{st}=\sigma\cdot R(a,L)[C(a)+\beta\Gamma(2-a)\cdot\sin(\pi(a-1)/2)]\cdot \qquad (28)$$

For $a$=1 (Cauchy): sin(0)=0→ $V_{st}=\sigma L$. (The asymmetry is not important, since $a$=1 is the limit). For $a$=2: sin(π/2)=1→ $V_{st}=\sigma \ln L$ (C(2)+ βΓ(0)). Here the regularization of the Γ-function comes into effect, leading to classical diffusion.

Calculating the "worst-case scenario" with maximum asymmetry (β=1) shows that the real safety threat is even higher than predicted by the symmetric model, since fluctuations in the reactor are directed predominantly "upward" (power surges).

Recalculation of the risk table for the "worst-case scenario" (β=1). For β=1, the stochastic contribution $V_{st}$ is amplified by the sinusoidal addition. The values of $\langle T\rangle(a)$, calculated using formula (26), which, unlike (28), uses $L^{\delta(a)}$ rather than R($a$,L), are compared for β=0 and β=1 in Table 1.

$$\langle T\rangle(a)=\Phi_{crit}/(v+V_{st}), \qquad V_{st}=\sigma L^{\delta(a)}\,[C(a)+\beta\Gamma(2-a)\sin(\pi(a-1)/2)] \tag{29}$$

Table 1.

| a | 3 | 2 | 1.5 | 1.1 | 1 |
|---|---|---|---|---|---|
| $\langle T\rangle(a)_{\beta=0}$ | 0.005 | 0.921 | 0.291 | 0.101 | 0.078 |
| $\langle T\rangle(a)_{\beta=1}$ | 0.005 | $1.999\cdot10^{-4}$ | 0.152 | 0.088 | 0.078 |

If you plot the graph, you will see two curves that merge at the points $a$=1 and $a$=3, but diverge maximally in the interval 1.5<$a$<2.5. At the point $a$=3 (Gaussian), the curves coincide. The β asymmetry disappears, as the distribution becomes normal. In the region $a$≈1.5 (the maximum gap), the influence of β=1 is most destructive. The safety time is reduced by 31% in addition to the stochastic effect. This is the zone where "Lévy flights" upwards occur frequently and have a huge amplitude. At the point $a$=1 (Cauchy), the curves converge again. In the Cauchy limit ($a$=1), the system scale L itself dominates, and the contribution of the asymmetry "saturates" (the sine from the formula vanishes).

The inclusion of the skewness parameter β=1 reveals a stochastic synergy: the combination of heavy-tailed fluctuations and positive power asymmetry leads to an additional 30% depletion of safety margins in the Lévy regime (1.5<$a$<2.0). While the Gaussian limit ($a$=3) remains invariant to skewness, the intermediate scaling region exhibits a "worst-case" scenario where the time-to-limit <T> falls below the critical 1- second threshold for standard protection systems. This highlights the necessity of accounting for the directional nature of power excursions in boiling-induced noise analysis.

Conceptual conclusion: "Stochastic Symmetry Breaking." Introducing the parameter $\beta$ into the time reserve model reveals the effect of stochastic symmetry breaking. In the classical Gaussian limit ($a$≥3), the system is symmetric: positive and negative power fluctuations are equally probable and cancel each other out. However, when transitioning to the Lévy regime ($a$<2), nonlinear connectivity in the core (the interaction of boiling and the neutron field) generates a preferred direction of probability transfer.

Mathematically, this is expressed by the fact that the asymmetry term $\beta$sin(…) becomes dominant in the critical region 1.5<$a$<2. Physically, this means that structural degradation (heavier tails) is inevitably accompanied by a dynamic imbalance: the system not only becomes noisier, it begins to "prefer" emergency acceleration trajectories. This transforms the parameter $\beta$ from a statistical formality into a critical indicator of directional reactor instability.

## Appendix C: stochastic term $V_{st}$

The stochastic term $V_{st}$, which determines the rate of approach to the threshold, is written in symmetrized form: $V_{st}(a, \beta, L) = \sigma\cdot\mathscr{R}(a, L)\cdot\Psi(a, \beta)$, where the structural risk factor $\Psi(a, \beta)$ combines the geometry of the distribution and its asymmetry equal to (17), (28):

$$\Psi(a, \beta) = \Gamma(2-a)\,[|\cos(\pi(a-1)/2)| + \beta\sin(\pi(a-1)/2)]$$

The gamma function Γ(2−$a$) in this connection plays the role of a "fluctuation intensity regulator". The scaling normalization Γ(2−$a$) is a natural weighting factor in the transition from Lévy jumps to continuous drift. It "tunes" the amplitude of the stochastic contribution depending on the tail heaviness. In the Cauchy regime ($a$→1) and Γ(1)=1. The coefficient becomes neutral, transferring control to the linear scaling L. This corresponds to the regime of "pure" ballistic flybys. In the Gaussian regime ($a$→2), the argument tends to zero (2−a→0), and the Γ-function exhibits a singularity (Γ(0)~1/x). However, in our formula, it is multiplied by the cosine and sine, and the cosine also tends to zero at this point. An "uncertainty unfolding" ∞0 occurs, which yields a finite limit of $\pi/2$≈1.57. The factor Γ(2−$a$) accelerates the stochastic behavior as the percolation critical point is approached, ensuring a correct transition to classical diffusion.

The combined stochastic factor $\Psi(a, \beta)$ ensures analytical continuity across all scaling regimes. While $\Gamma(2-a)$ formally diverges at the Gaussian limit ($a \to 2$), the simultaneous vanishing of the trigonometric terms leads to a finite limit: $a \to 2$, $\lim_{a\to 2}\Psi(a, \beta)=(\pi/2)(1+\beta\cdot 0)=\pi/2\approx 1.57$.

This confirms that the contribution of the skewness parameter $\beta$ is physically suppressed in the Gaussian regime, consistent with the symmetry requirements of the Central Limit Theorem. Conversely, in the Lévy regime ($1<a<2$), the term $\beta\sin(\pi(a-1)/2)$ acts as a directional amplifier, significantly reducing the safety margin for positive power excursions.

A relation is obtained that does not vanish at $a=1$ (due to $\mathscr{R}$), does not explode at $a=2$ (due to the mutual cancellation of Γ and trigonometry), and takes into account the asymmetry of $\beta$ only where it physically exists (in the Levy regime).

We will summarize all the calculations in a final table of constants (Table 2) and add a conceptual conclusion about "spontaneous breaking of stochastic symmetry".

The calculation is performed using the formula: $\Psi(a, \beta)=\Gamma(2-a)[|\cos(\pi(a-1)/2)|+\beta\sin(\pi(a-1)/2)]$, $\beta=1$ (maximum asymmetry of power surges).

Table 2.

| **Mode (a)** | $\mathbf{\Gamma(2-a)}$ | **Trigonometric block ($\beta$=1)** | $\mathbf{\Psi(a, 1)}$ | **Scaling $\mathscr{R}$(a,25)** | **Result $V_{st}(\sigma=0.002)$** |
|---|---|---|---|---|---|
| **1.0 (Cauchy)** | 1.00 | 1.0 +0.0=1.0 | **1.00** | 24.0 | **0.048** |
| **1.5 (Lévy)** | √π≈1.77 | 0.707 + 0.707= 1.41 | **2.51** | 8.0 | **0.040** |
| **2.0 (DP)** | →∝ | →0 | **1.57 (π/2)** | 3.2(lnL) | **0.010** |
| **3.0 (Gauss)** | -1.00 | 0.0-1.00=-1 | **1.00** | 0.96≈1 | **0.002** |

*Note: At the point a=2, the uncertainty ∝0 is resolved according to L'Hôpital's rule, yielding the constant π/2. At the point a=3, the modulus and signs of the Γ-function in the theory of stable laws are traditionally renormalized to unity.*

## Appendix D: Flowchart of the Online Monitoring Algorithm (CPS)

The following logic scheme is proposed for implementation in software control and protection systems (CPS): Block 1, Data Acquisition. Input: Signal from DPD (intra-band sensors), frequency 100 Hz. Output: Time series δP(t) for the last 20 seconds. Block 2, Spectral-fractal analysis (Processing). Step A: Trend removal (DFA detrending). Step B: Estimation of the Hurst exponent H. Step C: Estimation of the asymmetry β by calculating high-order moments or logarithmic moments. Block 3: Prediction time calculation. Input: Calculated $\alpha=a-1=1/H$, β, current intensity σ. Action: Solve the Lundberg equation with the regularizer $\mathscr{R}$(a,L). Result: Predicted time $\langle T\rangle_{pred}$. Block 4: Decision Making. If $\langle T\rangle_{pred} > \langle T\rangle_{safe}$: "Normal" mode. If $\langle T\rangle_{limit} < \langle T\rangle_{pred} < \langle T\rangle_{safe}$: Warning mode (shift of $\Phi_{crit}$ settings). If $\langle T\rangle_{pred} \leq \langle T\rangle_{limit}$: Emergency protection mode.